\documentclass[aps,prd, nofootinbib, notitlepage,onecolumn]{revtex4-1}
\usepackage{bm}
\usepackage{latexsym}
\usepackage{amsmath,amsfonts,amssymb}
\usepackage{graphicx,epsfig}
\usepackage{psfrag}
\usepackage{amsthm}
\usepackage{color}
\usepackage{braket}
\usepackage{indent first}
\usepackage[normalem]{ulem}
\usepackage{fancyhdr}
\usepackage{hyperref}
\usepackage{environ}
\usepackage{mathtools}
\usepackage{float}
\usepackage{soul}
\usepackage{bbold}
\newcommand{\be}{\begin{equation}}
\newcommand{\ee}{\end{equation}}
\newcommand{\bes}{\begin{equation*}}
\newcommand{\ees}{\end{equation*}}
\newcommand{\rarrow}{\rightarrow}
\newcommand{\bea}{\begin{eqnarray}}
\newcommand{\eea}{\end{eqnarray}}
\newcommand{\bseq}{\begin{subequations}}
\newcommand{\eseq}{\end{subequations}}

\NewEnviron{eqn}{
\begin{align}
\begin{split}
  \BODY
\end{split}
\end{align}
}

\NewEnviron{eqn*}{
\begin{align*}
\begin{split}
  \BODY
\end{split}
\end{align*}
}

\begin{document}
%%%%%%%%%%%%%%%%%%%%%%%%%%%%%%%%%%%%%
\title{Discriminating quantum gravity models by gravitational decoherence}
%%%%%%%%%%%%%%%%%%%%%%%%%%%%%
%
%
%
%
%
\author{Eissa Al-Nasrallah}
\email{eissa.alnasralla@grad.ku.edu.kw}
\affiliation{Department of Physics, Kuwait University, P.O. Box 5969, Safat 13060, Kuwait}
\affiliation{Energy and Building Research Center, Kuwait Institute for Scientific Research, P.O. Box 24885, Safat 13109, Kuwait}

\author{Saurya Das} 
\email{saurya.das@uleth.ca}
\affiliation{Theoretical Physics Group and Quantum Alberta, Department of Physics and Astronomy,
University of Lethbridge,
4401 University Drive, Lethbridge,
Alberta, T1K 3M4, Canada}

\author{Fabrizio Illuminati} \email{filluminati@unisa.it}
\affiliation{Dipartimento di Ingegneria Industriale, Universita degli Studi di  Salerno, Via Giovanni Paolo II, 132 - 84084 Fisciano, Salerno, Italy \\ INFN, Sezione di Napoli, Gruppo collegato di Salerno, Italy}

\author{Luciano Petruzziello} \email{lupertruzziello@unisa.it}
\affiliation{Dipartimento di Ingegneria Industriale, Universita degli Studi di  Salerno, Via Giovanni Paolo II, 132 - 84084 Fisciano, Salerno, Italy \\ INFN, Sezione di Napoli, Gruppo collegato di Salerno, Italy \\ Institut f\"ur Theoretische Physik, Albert-Einstein-Allee 11, Universit\"at Ulm, 89069 Ulm, Germany}

\author{Elias C. Vagenas} 
\email{elias.vagenas@ku.edu.kw}
\affiliation{Theoretical Physics Group, Department of Physics, Kuwait University, P.O. Box 5969, Safat 13060, Kuwait}

%%%%%%%%%%%%%%%%%%%%%%%%%%%%%%%%%%%%%%%%%%%%%%%%%%%%%%
\begin{abstract}
%%%%%%%%%%%%%%%%%%%%%%%%%%%%%%%%%%%%%%%%%%%%%%%%%%%%%%
\par\noindent
Several phenomenological approaches to quantum gravity predict the existence of a minimal measurable length and/or a maximum measurable momentum near the Planck scale. When embedded into the framework of quantum mechanics, such constraints induce a modification of the canonical commutation relations and thus a generalization of the Heisenberg uncertainty relations, commonly referred to as generalized uncertainty principle (GUP).
Different models of quantum gravity imply different forms of the GUP. For instance, in the framework of string theory the GUP is quadratic in the momentum operator, while in the context of doubly special relativity it includes an additional linear dependence. Among the possible physical consequences, it was recently shown that the quadratic GUP induces a universal decoherence mechanism, provided one assumes a foamy structure of quantum spacetime close to the Planck length. 
Along this line, in the present work we investigate the gravitational decoherence associated to the linear-quadratic GUP and we compare it with the one associated to the quadratic GUP. We find that, despite their similarities, the two generalizations of the Heisenberg uncertainty principle yield decoherence times that are completely uncorrelated and significantly distinct.
Motivated by this result, we introduce a theoretical and experimental scheme based on cavity optomechanics to measure the different time evolution of nonlocal quantum correlations corresponding to the two aforementioned decoherence mechanisms. We find that the deviation between the two predictions occurs on time scales that are macroscopic and thus potentially amenable to experimental verification. This scenario  provides a possible setting to discriminate between different forms of the GUP and therefore different models of quantum gravity.
\end{abstract}
\maketitle
%
%%%%%%%%%%%%%%%%%%%%%%%%%%%%%%%%%%%%%%%%%%%%%%%%%%%%%%
\section{Introduction}
%%%%%%%%%%%%%%%%%%%%%%%%%%%%%%%%%%%%%%%%%%%%%%%%%%%%%%
%
%
\par\noindent
Quantum decoherence has been an active area of research for more than 50 years. It was originally proposed by Zeh as an explanation for the quantum-to-classical transition \cite{Zeh:1970zz}. The conundrum of quantum systems behaving classically at the macroscopic scale has been a source of various interpretations and intense discussions. The phenomenon of decoherence attempts to solve this issue by taking into account the unavoidable entanglement shared between any quantum system and its environment. A similar entanglement affects the set of possible observable states of the system in a measurement and in turn gives rise to a net loss of quantum information. In other words, the system tends to lose all the characteristic quantum features (such as for instance the superposition of states which are predicted quantum mechanically and yet unobserved on a macroscopic scale, and in some circumstances even prohibited \cite{Zurek:2003zz, Schlosshauer:2019ewh,Schlosshauer:2003zy,Hornberger, Anastopoulos:2000hg}).\\ 
\indent
In the quest for seeking a faithful and rigorous representation of natural phenomena, it has been observed that our most successful theories, namely quantum field theory (QFT) and general relativity, encounter non-trivial adversities that prevent them from providing an accurate description of reality when approaching the Planck scale. This stimulated the search for a consistent theory of quantum gravity that has led to the emergence of several promising candidates; among them, it is worth mentioning string theory, loop quantum gravity, causal dynamical triangulations, modified dispersion relations and doubly special relativity (DSR). All these models predict a minimal measurable length which is typically proportional to the Planck length $\mathcal{O}(\ell_{p})$, where $\ell_{p}\sim10^{-35}$ m. Such a constraint on the minimal measurable length necessarily requires a modification of the Heisenberg uncertainty principle (HUP), which goes by the name of generalized uncertainty principle (GUP) (see Refs. \cite{Kempf:1994su,noncom2,noncom3,bis,quar,ses,ong,set,ot,dieci,qft,qft4,plenio2,ourplb,ourprd,Das:2009hs,Das:2010sj,rnd,rnd2} and therein for a comprehensive overview of the subject). The incompatibility between the absence of a fundamental limit to the spatial resolution as conveyed by the HUP and the presence of gravity has been originally put forward in the context of superstring scattering processes at Planckian energies~\cite{gener}. However, similar results can also be achieved in many other different scenarios, for instance when dealing with gedanken experiments involving micro~\cite{gener2} and macro~\cite{gener3} black holes, thus further corroborating the intrinsic relevance of gravitationally-induced deformations of the standard quantum mechanical scheme.

Understanding decoherence in the framework of quantum gravity is thought to bring an insight on the nature of quantum gravitational interactions and provide possibilities for their experimental tests. Decoherence stemming from general relativity and different models of quantum gravity has been thoroughly studied in the recent literature (see for instance Refs. \cite{Asprea:2019dok, Allali:2021puy, Anastopoulos:2013zya,Bonder:2015hja,Miki:2020hvg,Podolskiy:2015wna,Pang:2016foq,ellis,milburn,bargueno} and therein). Recently, two of the present authors have considered the deformed canonical commutation relations (DCCRs) implied by string theory \cite{Kempf:1994su,Kempf:1996fz,Das:2017qwp}, and have derived a Lindblad-type master equation valid for a generic quantum system when the deformation parameter is assumed to be stochastic \cite{Petruzziello:2020wkd}. The string-theory inspired DCCRs read
%
%
%%%%%%%%%%%%%%%%%% equation %%%%%%%%%%%%%%%%%%%%%%%%%%%
\begin{equation}
\label{QGUP_DCCR}
    [x_i,p_j] = i\hbar \left[ \delta_{ij} + \beta  \delta_{ij}  p^2+ 2\beta p_{i}p_{j} \right], \ \ \beta=\beta_{0}\frac{\ell_{p}^2}{\hbar^2},~
\end{equation}
with $\beta$ being the deformation parameter. The corresponding generalized uncertainty relations are
%
%
%%%%%%%%%%%%%%%%%% equation %%%%%%%%%%%%%%%%%%%%%%%%%%%
\begin{equation}
\label{QGUP}
    \Delta x \Delta p \geq \frac{\hbar}{2} \left (1+\beta \Delta p^2 \right)~.
\end{equation}
\par\noindent
This GUP is quadratic in the momentum operator and, when equipped with a fluctuating $\beta$, is the one that has been discussed in the context of quantum gravitational decoherence \cite{Petruzziello:2020wkd}. Here, we wish to analyze a different form of GUP, known as LQGUP, which includes terms that are both linear and quadratic in the momentum operator. Such a generalization is compatible with doubly special relativity and arises in various contexts \cite{Vagenas:2019wzd, Das:2008kaa, Das:2020ujn, Ali:2010yn, Ali:2009zq, Das:2020ujn, Ali:2010yn}. The standard formulation of the LQGUP is
%
%
%
%%%%%%%%%%%%%%%%%% equation %%%%%%%%%%%%%%%%%%%%%%%%%%%
\bea
\label{lqgup}
\Delta x \Delta p 
&&\geq\frac{\hbar}{2}\left[1+\left(\frac{\alpha}{\sqrt{\langle p^2 \rangle}}+4\alpha^2\right)\Delta p^2+4\alpha^2\langle p \rangle^2-2\alpha\sqrt{\langle p^2 \rangle}\right]\nonumber\\
&&\geq \frac{\hbar}{2}\left[1-2\alpha \langle p \rangle +4 \alpha^2 \langle p^2 \rangle \right]~.
\eea
\par\noindent
 In this case, the DCCRs are
%
%
%%%%%%%%%%%%%%%%%% equation %%%%%%%%%%%%%%%%%%%%%%%%%%%
\begin{equation}
\label{canonical_gup}
    [x_i,p_j] = i \hbar \left[\delta_{ij} - \alpha \left(p\delta_{ij} + \frac{p_ip_j}{p} \right) + \alpha^2 (p^2 \delta_{ij} + 3p_ip_j)\right]~.
\end{equation}
The operators $x_i$ and $p_i$ are defined as
%%%%%%%%%%%%%%%%%% equation %%%%%%%%%%%%%%%%%%%%%%%%%%%
\begin{equation}
\label{definitions_x_p}
    x_i = x_{0i}, \  p_i = p_{0i}(1-\alpha p_{0} + 2\alpha^2p_{0}^2)~,\\
\end{equation}
where the operators $x_{0i}$ and $p_{0i}$ satisfy the standard CCRs
%
%%%%%%%%%%%%%%%%%% equation %%%%%%%%%%%%%%%%%%%%%%%%%%%
\begin{equation}
[x_{0i},p_{0j}] = i\hbar \delta _{ij},
\end{equation}
and $[x_i,x_j] = 0, [p_i,p_j] = 0$, $p^2 = \sum_{j=1}^3 p_jp_j, p_{0}^2 = \sum _{j=1}^3 p_{0j}p_{0j}$.
The deformation parameter is $\alpha = {\alpha_{0}}/{(M_{P}c)} = {\alpha_{0} \ell_{p}}/{\hbar}$, with $\alpha_{0}$ assumed of order unity \cite{Ali:2010yn}. 

In this paper, by promoting $\alpha$ to a fluctuating quantity encoding the foamy structure of spacetime at the Planck scale, we derive a quantum gravitational decoherence mechanism associated to the LQGUP and pinpoint the main differences with the one corresponding to the quadratic GUP \cite{Petruzziello:2020wkd}. This provides a tool to distinguish and discriminate between different phenomenological approaches to quantum gravity, such as string theory and doubly special relativity. We derive and compare the decoherence times as well as the decoherent dynamics of quantum correlations predicted by the two models. Specifically, we introduce an experimental scheme capable of detecting the tiny discrepancies between the two gravitational decoherence mechanisms. Such a scheme is based on cavity optomechanics and measures the time evolution of Clauser-Horne-Shimony-Holt (CHSH) nonlocal correlations, which represent the central quantities appearing in the Bell inequalities \cite{Pfister:2015sna}. By studying the deviation among the two predicted behaviors with suitable choices of the experimental parameters, we find that the discrepancy occurs at a macroscopic and thus experimentally accessible scale. This scenario offers the opportunity to discriminate  different approaches to quantum gravity via their low-energy phenomenology in terms of gravitational decoherence.

The paper is structured as follows:
in Section II we derive the Lindblad-type master equation in the context of LQGUP, whereas in Section III we obtain the entropy variation and the decoherence time due to the LQGUP-modified Lindblad master equation. 
In Section IV we show how to discriminate experimentally the two models of quantum gravitational decoherence associated to the GUP and to the LQGUP.
Finally, in Section V we discuss our results and possible future perspectives.
%
%
%%%%%%%%%%%%%%%%%%%%%%%%%%%%%%%%%%%%%%%%%%%%%%%%%%%%%
\section{LQGUP-induced master equation}
%%%%%%%%%%%%%%%%%%%%%%%%%%%%%%%%%%%%%%%%%%%%%%%%%%%%%
%
%
%
%
\par\noindent
Following the analysis of Ref. \cite{Petruzziello:2020wkd}, we derive the Lindblad-type master equation in the context of LQGUP. 
First, we note that we can start without loss of generality from the canonical Schr\"odinger equation 
%
%%%%%%%%%%%%%%%%%% equation %%%%%%%%%%%%%%%%%%%%%%%%%%%
\begin{equation}
\label{Schrodinger eqn}
 i \hbar \partial_t |\psi\rangle  = H|\psi\rangle = \left(\frac{p^2}{2m} + V \right) |\psi\rangle    
\end{equation}
and use the definition of the GUP-modified momentum as given by Eq. \eqref{definitions_x_p} to get
%
%
%%%%%%%%%%%%%%%%%% equation %%%%%%%%%%%%%%%%%%%%%%%%%%%
\begin{equation}
\label{explicit_GUP_Schrodinger_eqn}
    i \hbar \partial_t |\psi\rangle = \left(p_{0}^2\frac{\left(1-\alpha p_{0} + 2\alpha^2p_{0}^2\right)^2}{2m} + V\right)|\psi\rangle=\left(p_{0}^2\frac{\left(1-2\alpha p_{0} + 5\alpha^2p_{0}^2 - 4 \alpha^3 p_{0}^3 + 4\alpha^4 p_{0}^4\right)}{2m} + V\right)|\psi\rangle 
\end{equation}
%
%
%%%%%%%%%%%%%%%%%% equation %%%%%%%%%%%%%%%%%%%%%%%%%%%
and thus
\begin{equation}
\label{full_LQGUP}
    i \hbar \partial_t |\psi\rangle = \left(\frac{p_{0}^2}{2m} + V -\alpha\frac{p_{0}^3}{m}  + 5\alpha^2\frac{p_{0}^4}{2m} - 2 \alpha^3\frac{p_{0}^5}{m}  + 2 \alpha^4\frac{p_{0}^6}{m} \right)|\psi\rangle~.
\end{equation} 
We observe that several orders of the deformation parameter appear in the perturbation of the standard Hamiltonian. For our purposes, we will consider a first-order treatment of the factor $\alpha$. However, for the sake of completeness, we show in Appendix \ref{appendix_A} that considerations involving second-order corrections produce the same outcome.
\newline
Hence, if we account for a first-order treatment only of the GUP parameter $\alpha$ and ignore factors that go as $\mathcal{O}(\alpha^2)$, then the Schr\"odinger equation becomes
%
%
%%%%%%%%%%%%%%%%%% equation %%%%%%%%%%%%%%%%%%%%%%%%%%%
\begin{equation}
\label{concise_Schrodinger_eqn}
i \hbar \partial_t |\psi\rangle = H|\psi\rangle = \left(H_0 + H_{1} + \mathcal{O}(\alpha^2) \right) |\psi\rangle~, 
\end{equation}
where 
%
%%%%%%%%%%%%%%%%%% equation %%%%%%%%%%%%%%%%%%%%%%%%%%%
\begin{equation}
\label{hamiltonian_identities}
    H_0 = \frac{p_{0}^2}{2m} + V, \ \  H_{1} = - 2\alpha\frac{p_{0}^3}{2m} ~.
\end{equation}
\\
Now, by introducing the density matrix  
\begin{equation}
\label{densitymatrix}
\varrho=|\psi\rangle\langle\psi|~,
\end{equation}
we can cast the Liouville-von Neumann equation as follows
%
%%%%%%%%%%%%%%%%%% equation %%%%%%%%%%%%%%%%%%%%%%%%%%%
\begin{equation}
\label{liouville_von_neumann}
    \partial_t\varrho = -\frac{i}{\hbar}[H_{0} + H_{1}, \varrho]~.
\end{equation}
Next, to solve the equation for the density matrix we shift our analysis to the interaction picture, where the state takes the form
%
%
%%%%%%%%%%%%%%%%%% equation %%%%%%%%%%%%%%%%%%%%%%%%%%%
\begin{equation}
\label{interaction_pic}
   |\psi\rangle = e^{-\frac{iH_{0}t}{\hbar}}|\Tilde{\psi}\rangle~. 
\end{equation}
According to this representation, we can rewrite Schr\"odinger equation as
%
%
%%%%%%%%%%%%%%%%%% equation %%%%%%%%%%%%%%%%%%%%%%%%%%%
\begin{equation}
\label{unitary_transformed_hamiltonian}
    i \hbar \partial_t|\Tilde{\psi}\rangle = \Tilde{H_{1}} |\Tilde{\psi}\rangle \ \text{where}  \ \ \Tilde{H_{1}} = e^{\frac{iH_{0}t}{\hbar}} H_{1} e^{-\frac{iH_{0}t}{\hbar}}~,
\end{equation}
and substituting back in the Liouville-von Neumann equation, i.e., Eq. \eqref{liouville_von_neumann}, we get
%%%%%%%%%%%%%%%%%% equation %%%%%%%%%%%%%%%%%%%%%%%%%%%
\begin{equation}
\label{perturbed_liouville_von_neumann}
    \partial_t\tilde{\varrho}(t) =  -\frac{i}{\hbar}[\tilde{H_{1}}(t),\tilde{\varrho}(t)]~.
\end{equation}
A formal solution of the above equation can be achieved by integrating both sides over the interval $[0,t]$, which thus yields
%
%
%%%%%%%%%%%%%%%%%% equation %%%%%%%%%%%%%%%%%%%%%%%%%%%
\begin{equation}
\label{integrated_liouville}
    \int_{0}^{t} \partial_t\tilde{\varrho}(t')dt' =  \int_{0}^{t}-\frac{i}{\hbar}[\tilde{H_{1}}(t'),\tilde{\varrho}(t')]dt'
\end{equation}
and results in
%
%
%%%%%%%%%%%%%%%%%% equation %%%%%%%%%%%%%%%%%%%%%%%%%%%
\begin{equation}
\label{integrated_perturbed_liouville_von_neumann}
\tilde{\varrho}(t) = \tilde{\varrho}(0) -\frac{i}{\hbar}\int_{0}^{t}[\tilde{H_{1}}(t'),\tilde{\varrho}(t')]dt'.
\end{equation}
Subsequently, Eq. \eqref{integrated_perturbed_liouville_von_neumann} is inserted into the r.h.s. of Eq. \eqref{perturbed_liouville_von_neumann}, thereby giving 
%
%
%%%%%%%%%%%%%%%%%% equation %%%%%%%%%%%%%%%%%%%%%%%%%%%
\begin{equation}
\label{liouville_solution}
  \partial_t\tilde{\varrho}(t) = -\frac{i}{\hbar}[\tilde{H_{1}}(t),\tilde{\varrho}(0)] -\frac{1}{\hbar^2} \int_{0}^{t}[\tilde{H_{1}}(t),[\tilde{H_{1}}(t'),\tilde{\varrho}(t')]]dt'~.  
\end{equation}
The next step consists in relying on the Born-Markov approximation to let the density matrix of the r.h.s. of the above equation depend on $t$ rather than $t'$, i.e., 
%
%
%%%%%%%%%%%%%%%%%% equation %%%%%%%%%%%%%%%%%%%%%%%%%%%
\begin{equation}
\label{born_markov_approximation}
    \partial_t\tilde{\varrho}(t) = -\frac{i}{\hbar}[\tilde{H_{1}}(t),\tilde{\varrho}(0)] -\frac{1}{\hbar^2} \int_{0}^{t}[\tilde{H_{1}}(t),[\tilde{H_{1}}(t'),\tilde{\varrho}(t)]]dt'.
\end{equation}
This is a common procedure in the derivation of Lindblad-type master equations for weak interactions between the system and reservoir \cite{Petruzziello:2020wkd,Xu:2020pzr,Kolovsky_2020}. However, we note that it has been argued that this approximation may not be valid for certain regimes, such as composite systems in strong coupling regime \cite{Nakatani2010} and low temperature systems \cite{Vadimov2021}.\\
\indent
At this point, we observe that the GUP based on theories such as Loop Quantum Gravity and DSR should contain the information of space-time  fluctuations near the Planck scale, which is a typical by-product of the aforementioned quantum gravity models. To comply with the existence of a fluctuating minimum length scale, one can think of encoding a similar feature in the present analysis by demanding a fluctuating deformation parameter, which requires a stochastic treatment of the problem. By averaging over the fluctuations of the GUP parameter $\alpha$ we can allow for the development of the master equation from the mean stochastic Liouville-von Neumann equation. Therefore, we have to regard $\alpha$ as a random variable; specifically, for the sake of simplicity we can interpret it as being a Gaussian white noise with fixed mean and sharp auto-correlation. Inspired by the considerations contained in Ref. \cite{Petruzziello:2020wkd}, we can then impose that the dimensionless GUP parameter $\alpha_0$ is given by
%
%
%%%%%%%%%%%%%%%%%% equation %%%%%%%%%%%%%%%%%%%%%%%%%%%
\begin{equation}
\label{average_alpha}
    \alpha_{0} =\sqrt{t_{p}}\chi(t), \ \ \langle\chi(t)\rangle= \Bar{\alpha}_{0},\ \  \langle\chi(t)\chi(t')\rangle = \delta(t-t')~,
\end{equation}
with $t_p$ being the Planck time.
In order to be in accordance with the standard literature, we can assume $\alpha_0$ to be of order unity on average. For this purpose, if we average over its fluctuations we can take $\langle\alpha_0\rangle=\sqrt{t_{p}}\langle\chi(t)\rangle=1$ and consequently define the fixed mean as $\Bar{\alpha}_{0} = {1}/{\sqrt{t_{p}}}$. Alternatively, we can follow a more conservative approach and impose the mean value of $\alpha_0$ to not exceed the experimental bound available for the deformation parameter, that is $\langle\alpha_0\rangle\simeq 10^{15}$. In any case, we will show that the arbitrary choice made to fix $\bar{\alpha}_0$ will not play a relevant role in the upcoming investigation, as the important quantity is represented by the auto-correlation and not the mean value.\\
Now, by averaging over the fluctuations and introducing the shorthand notation $\langle \varrho \rangle = \rho$, the derivative will take the form
%
%
%%%%%%%%%%%%%%%%%% equation %%%%%%%%%%%%%%%%%%%%%%%%%%%
\begin{equation}
\label{fluctuations_average}
    \partial_t \langle\tilde{\varrho}\rangle = \partial_t \tilde{\rho}(t),
\end{equation}
and Eq. \eqref{born_markov_approximation}  can be rewritten as
%
%
%%%%%%%%%%%%%%%%%% equation %%%%%%%%%%%%%%%%%%%%%%%%%%%
\begin{equation}
\label{average_liouvile_von_neumann}
\partial_t \tilde{\rho}(t) =  \left \langle -\frac{i}{\hbar}\left[\tilde{H}_{1}(t),\tilde{\varrho}(0)\right] \right \rangle - \frac{1}{\hbar^2} \int_{0}^{t}\left \langle \left[\tilde{H}_{1}(t),\left[\tilde{H}_{1}(t'),\tilde{\varrho}(t)\right]\right] \right \rangle dt'.
\end{equation}
The first term on the r.h.s. of Eq. \eqref{average_liouvile_von_neumann} vanishes since $\tilde{\varrho}(0)$ is constant. For the second term, in order to further manipulate the factor inside the integral, we recall the definition of $H_{1}$ in Eq. \eqref{hamiltonian_identities} to substitute it and obtain
%
%
%%%%%%%%%%%%%%%%%% equation %%%%%%%%%%%%%%%%%%%%%%%%%%%
\begin{equation}
\label{prefinal_integral_version}
\partial_t \tilde{\rho}(t) = - \frac{4}{\hbar^2} \int_{0}^{t}\left \langle \alpha(t) \alpha(t') \right \rangle \left[\frac{\tilde{p}^3_0(t)}{2m} ,\left[\frac{\tilde{p}^3_0(t')}{2m} ,\tilde{\rho}(t)\right]\right]  dt'.
\end{equation}
In terms of  $\alpha={\alpha_{0} \ell_{p}}/{\hbar}$, we rewrite the above equation as
%
%
%%%%%%%%%%%%%%%%%% equation %%%%%%%%%%%%%%%%%%%%%%%%%%%
\begin{equation}
\label{intermediate}
\partial_t \tilde{\rho}(t) = - \frac{4\ell_{p}^2}{\hbar^4} \int_{0}^{t}\left \langle \alpha_0(t) \alpha_0(t') \right \rangle \left[\frac{\tilde{p}^3_0(t)}{2m},\left[\frac{\tilde{p}^3_0(t')}{2m},\tilde{\rho}(t)\right]\right]  dt'~.
\end{equation}
Now since $\alpha_0$ is a Gaussian white noise as explained in Eq. (\ref{average_alpha}), the integral can be easily evaluated (See Appendix \ref{appendix_A}), thereby yielding
%
%%%%%%%%%%%%%%%%%% equation %%%%%%%%%%%%%%%%%%%%%%%%%%%
\begin{equation}
\label{master_eqn_momentum_representation}
  \partial_t \tilde{\rho}(t) =   -\sigma \left [\frac{\tilde{p}^3_0(t)}{2m}, \left[\frac{\tilde{p}^3_0(t)}{2m},\tilde{\rho}(t) \right]\right ],
\end{equation}
where $\sigma =  {4t_{p}\ell_{p}^2}/{\hbar^4}$.
\par\noindent
At this point, we can recover the density matrix in the Schr\"odinger representation by means of the following transformation:
%
%
%%%%%%%%%%%%%%%%%% equation %%%%%%%%%%%%%%%%%%%%%%%%%%%
\begin{equation}
\label{density_matrix_Schrodinger_pic}
    \rho(t) = e^{-\frac{iH_{0}t}{\hbar}} \tilde{\rho}(t) e^{\frac{iH_{0}t}{\hbar}}~,
\end{equation}
so that we can go back to the Liouville-von Neumann equation given by Eq. \eqref{liouville_von_neumann} with the dissipator term that depends on the characteristic features of the stochastic LQGUP model treated so far. In particular, by requiring the system to not be subject to any external potential, i.e., $V=0$, we are left with
%
%
%%%%%%%%%%%%%%%%%% equation %%%%%%%%%%%%%%%%%%%%%%%%%%%
\begin{equation}
\label{lindbladian}
    \partial_t \rho(t) = -\frac{i}{\hbar} \left[\frac{p_0^2}{2m},\rho(t)\right]  - \sigma  \left [\frac{p^3_0}{2m},\left[\frac{p^3_0}{2m},\rho(t)\right]\right]~.
\end{equation}
\par\noindent
By employing the definitions in Eq. \eqref{hamiltonian_identities}, the above equation can be rewritten in terms of Hamiltonians  as 
%
%
%%%%%%%%%%%%%%%%%% equation %%%%%%%%%%%%%%%%%%%%%%%%%%%
\begin{equation}
\label{lindbladian_in_Hamiltonians}
    \partial_t \rho(t) = -\frac{i}{\hbar} \left[H_0,\rho(t)\right]  - 2m\sigma  \left [H^{3/2}_0,\left[H^{3/2}_0,\rho(t)\right]\right]~.
\end{equation}
%
%
%\par\noindent
Equations (\ref{lindbladian}) and (\ref{lindbladian_in_Hamiltonians}) represent the LQGUP-modified Lindblad-type master equation, where the dissipator is identified with the second term on the r.h.s.. Clearly, in the limit $\sigma\approx 0$ we recover the standard unitary dynamics predicted by the unmodified Liouville-von Neumann equation.
However, it is important to stress that, should the ansatz in Eq.~\eqref{average_alpha} not be considered and the LQGUP approach be assumed without further requirements, the source of decoherence in Eq.~\eqref{lindbladian_in_Hamiltonians} would not appear in the master equation. Indeed, it is well-known that, generally speaking, minimal-length-induced deformations of quantum mechanics give rise to a unitary time evolution~\cite{unita,unita2}.

Furthermore, it is worth remarking that the power of $H_0$ appearing in Eq. (\ref{lindbladian_in_Hamiltonians}) is different from the usual one that appears in other gravitational decoherence mechanisms, according to which the dependence on the free Hamiltonian is typically linear  \cite{ellis,milburn}. This unconventional behavior is shared also with the other work dealing with a GUP-induced decoherence process \cite{Petruzziello:2020wkd}, but interestingly the power of $H_0$ is different in the two scenarios. At variance with Eq. \eqref{lindbladian_in_Hamiltonians}, the counterpart of the LQGUP-modified Lindblad-type master equation for the quadratic GUP is given by
\begin{equation}
\label{lindbladian_in_Hamiltonians_GUP}
    \partial_t \rho(t) = -\frac{i}{\hbar} \left[H_0,\rho(t)\right]  - \sigma'  \left [H^{2}_0,\left[H^{2}_0,\rho(t)\right]\right]~, \qquad 
    \sigma'=\frac{16 m^2 \ell_p^4 t_p}{\hbar^6}\,.
\end{equation}
Such a difference indicates that the intrinsic characteristics of the emergent decoherence phenomenon strictly depend on the GUP model used for computations.
%
%
%%%%%%%%%%%%%%%%%%%%%%%%%%%%%%%%%%%%%%%%%%%%%%%%%%%%%
\section{Applications and Physical Results}
%%%%%%%%%%%%%%%%%%%%%%%%%%%%%%%%%%%%%%%%%%%%%%%%%%%%%
%
%
\par\noindent
In this Section, we look at the temporal evolution of an entropy quantifier, i.e., linear entropy (for which calculations can be carried out analytically), to analyze the irreversibility of the decoherence mechanism and provide a solution for the differential equation Eq. (\ref{lindbladian_in_Hamiltonians}). By means of this study, we are able to derive the decoherence time associated with the LQGUP, which foresees energy localization in momentum space as it occurs also in Ref. \cite{Petruzziello:2020wkd}.
%
%
%%%%%%%%%%%%%%%%%%%%%%%%%%%%%%%%%%%%%%%%%%%%%%%%%%%%%%
\subsection{Entropy Variation}
%%%%%%%%%%%%%%%%%%%%%%%%%%%%%%%%%%%%%%%%%%%%%%%%%%%%%%
%
%
%
%
\par\noindent
Linear entropy corresponds to the mixedness of the quantum state and is therefore associated with the quantum state purity. It is defined as
%
%
%%%%%%%%%%%%%%%%%% equation %%%%%%%%%%%%%%%%%%%%%%%%%%%
\begin{equation}
\label{entropy}
    S(t) = 1 - tr\left(\rho^2(t)\right),
\end{equation}
and we recall that a given state is pure if and only if the correlated density matrix is idempotent, namely $\rho^2=\rho$~.
Now, the time evolution of the linear entropy is obtained by differentiating Eq. \eqref{entropy}, that is
%
%
%%%%%%%%%%%%%%%%%% equation %%%%%%%%%%%%%%%%%%%%%%%%%%%
\begin{equation}
\label{linear_entropy_derivate}
    \partial_t  S(t) = \partial_t ( 1 - tr\left(\rho^2(t)\right)) = -2 tr \left( \rho(t) \partial_t\rho(t) \right).
\end{equation}
At this point, by resorting to the LQGUP-modified Lindblad-type master equation from the previous Section, namely Eq. \eqref{lindbladian}, we can substitute the factor $\partial_t\rho(t)$ to give
%
%
%%%%%%%%%%%%%%%%%% equation %%%%%%%%%%%%%%%%%%%%%%%%%%%
\begin{equation}
\label{Lindbladian_entropy}
    \partial_t  S = \frac{2i}{\hbar}  tr \left( \rho(t)\left[\frac{p_0^2}{2m},\rho(t)\right] \right) + 2 \  \sigma \ tr\left( \rho(t) \left[\frac{p^3_0}{2m},\left[\frac{p^3_0}{2m},\rho(t)\right]\right] \right).
\end{equation}
By manipulating the r.h.s of the above equation and using the cyclic property of the trace, one obtains
%
%
%%%%%%%%%%%%%%%%%% equation %%%%%%%%%%%%%%%%%%%%%%%%%%%
\begin{equation}
\label{entropy_with_traces}
\partial_t  S =  2 \  \sigma \ \left[ 2 \ tr \left(\rho \frac{p^6_0}{4m^2}\rho \right) - 2 \ tr \left(\rho \frac{p^3_0}{2m} \right)^2\right].
\end{equation}
If we now introduce operator $O$ defined as
%
%
%%%%%%%%%%%%%%%%%% equation %%%%%%%%%%%%%%%%%%%%%%%%%%%
\begin{equation}
\label{operator_o}
O = \left[ \frac{p^3_0}{2m}, \rho \right] , 
\end{equation}
then it can be shown that 
%
%
%%%%%%%%%%%%%%%%%% equation %%%%%%%%%%%%%%%%%%%%%%%%%%%
\begin{equation}
\label{operator_o_trace}
     \left[ 2 \ tr \left(\rho \frac{p^6_0}{4m^2} \rho \right)- 2 \ tr\left( \left(\rho \frac{p^3_0}{2m} \right)^2 \right) \right]= tr \left( O^{\dag} O\right),
\end{equation}
\par\noindent
where the result of the trace is always greater or equal to zero since the quantity $O^\dag O$ is by construction a positive semi-definite operator. Next, by going back to Eq. \eqref{entropy_with_traces}, we are left with
%
%%%%%%%%%%%%%%%%%% equation %%%%%%%%%%%%%%%%%%%%%%%%%%%
\begin{equation}
    \label{entropy_operator_o}
    \partial_t S = 2 \ \sigma \ tr \left( O^{\dag} O\right) \geq 0~,
\end{equation}
where $\sigma$ and $tr \left(O^{\dag}O \right)$ are positive quantities, which entails that the linear entropy is monotonically increasing with time and the initial quantum state tends to increase mixedness. Furthermore, this achievement clearly conveys the intrinsic irreversible nature of the studied process, as the entropy can never decrease with time.
%
%
%%%%%%%%%%%%%%%%%%%%%%%%%%%%%%%%%%%%%%%%%%%%%%%%%%%%%%%%%%%
\subsection{Decoherence Time}
%%%%%%%%%%%%%%%%%%%%%%%%%%%%%%%%%%%%%%%%%%%%%%%%%%%%%%%%%%%
%
%
%
\par
\begin{figure}[ht]
    \centering
    \includegraphics[width=17cm, height=10cm]{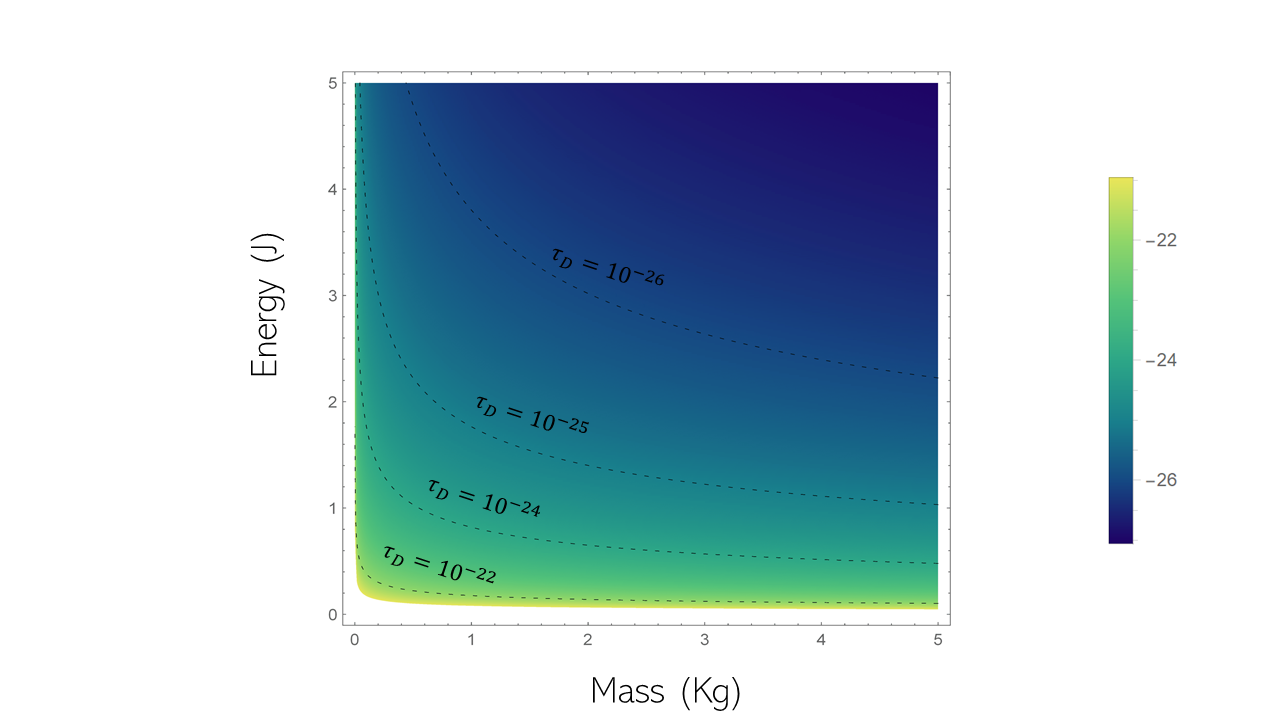}
    \caption{Plot of $\log_{10}(\tau_D)$ as a function of mass and energy.}
    \label{fig:my_label1}
\end{figure}

\noindent
We are now ready to calculate the decoherence time associated with the LQGUP-modified Lindblad-type master equation, i.e., Eq. \eqref{lindbladian_in_Hamiltonians}. To this aim, we point out that the computation is easier if performed in the momentum representation, where the elements of the density matrix are
%
%
%%%%%%%%%%%%%%%%%% equation %%%%%%%%%%%%%%%%%%%%%%%%%%%
\begin{equation}
\label{density_in_momentum_representation}
    \rho_{p,p'}(t) = \langle \bm{{\rm p}}|\rho(t)|\bm{{\rm p}}'\rangle ~,
\end{equation}
and consequently
%
%
%%%%%%%%%%%%%%%%%% equation %%%%%%%%%%%%%%%%%%%%%%%%%%%
\begin{equation}
\label{density_matrix_derivative}
    \partial_t \rho_{p,p'}(t) = \partial_t \langle \bm{{\rm p}}|\rho|\bm{{\rm p}}'\rangle = \langle \bm{{\rm p}}|\partial_t \rho|\bm{{\rm p}}'\rangle ~.
\end{equation}
Therefore, by using the master equation, i.e., Eq.  \eqref{lindbladian_in_Hamiltonians}, we get
%
%
%%%%%%%%%%%%%%%%%% equation %%%%%%%%%%%%%%%%%%%%%%%%%%%
\begin{equation}
\label{density_matrix_derivative_in_master_eqn}
\partial_t \rho_{p,p'}(t) = -\frac{i}{\hbar} \langle \bm{{\rm p}}|[{H}_0, {\rho}]|\bm{{\rm p}}'\rangle - 2m\sigma \langle \bm{{\rm p}}|\left[ {H}^{3/2}_0, [{H}^{3/2}_0,{\rho}]\right]|\bm{{\rm p}}'\rangle~.
\end{equation}
By expanding the commutators, one can easily check that 
%
%
%%%%%%%%%%%%%%%%%% equation %%%%%%%%%%%%%%%%%%%%%%%%%%%
\bea
\label{density_matrix_evolution_expanded_commutators}
\partial_t \rho_{p,p'}(t) &=& -\frac{i}{\hbar} \left( \langle \bm{{\rm p}}|{H}_0 {\rho}|\bm{{\rm p}}'\rangle - \langle \bm{{\rm p}}|{\rho} {H}_0|\bm{{\rm p}}'\rangle \right) \nonumber\\
&&- 2m\sigma \left( \langle \bm{{\rm p}}|{H}^3_0{\rho}|\bm{{\rm p}}'\rangle - \langle \bm{{\rm p}}|{H}^{3/2}_0 {\rho} {H}^{3/2}_0|\bm{{\rm p}}'\rangle - \langle \bm{{\rm p}}|{H}^{3/2}_0 {\rho} {H}^{3/2}_0|\bm{{\rm p}}'\rangle + \langle \bm{{\rm p}}|{\rho} {H}^3_0|\bm{{\rm p}}'\rangle \right).
\eea
In the momentum representation, we know that ${H}(p)|\bm{{\rm p}}\rangle = E(p)|\bm{{\rm p}}\rangle$. Moreover, by using the fact that the Hamiltonian is a Hermitian operator, we obtain
%
%
%%%%%%%%%%%%%%%%%% equation %%%%%%%%%%%%%%%%%%%%%%%%%%%
\bea
\label{density_matrix_evolution_in_energies}
\partial_t \rho_{p,p'}(t) &=& -\frac{i}{\hbar} \left( E(p)\langle \bm{{\rm p}}| {\rho}|\bm{{\rm p}}'\rangle - E(p')\langle \bm{{\rm p}}|{\rho}|\bm{{\rm p}}'\rangle \right)\nonumber \\
 &&- 2m\sigma \left( E^3(p)\langle \bm{{\rm p}}|{\rho}|\bm{{\rm p}}'\rangle - 2 E^{3/2}(p) E^{3/2}(p')\langle \bm{{\rm p}}|{\rho}|\bm{{\rm p}}'\rangle  + E^3(p')\langle \bm{{\rm p}}|{\rho} |\bm{{\rm p}}'\rangle \right). 
\eea
By writing the density matrix elements as in Eq. \eqref{density_in_momentum_representation}, the previous equation reads 
%
%
%%%%%%%%%%%%%%%%%% equation %%%%%%%%%%%%%%%%%%%%%%%%%%%
\begin{equation}
\label{density_matrix_evolution_in_energies_shortened}
    \partial_t \rho_{p,p'}(t) = -\frac{i}{\hbar} \left(E(p)  - E(p')\right) \rho_{p,p'} - 2m \sigma \left( E^3(p) - 2 E^{3/2}(p) \ E^{3/2}(p') +   E^3(p') \right)\rho_{p,p'} 
\end{equation}
and introducing the notation $\Delta E^{3/2} \equiv E^{3/2}(p) - E^{3/2}(p')$, we get
%
%
%%%%%%%%%%%%%%%%%% equation %%%%%%%%%%%%%%%%%%%%%%%%%%%
\begin{equation}
\label{energy_master_eqn}
    \partial_t \rho_{p,p'} = \left[ -\frac{i}{\hbar} \left(E(p)  - E(p')\right) - 2m \sigma (\Delta E^{3/2})^2 \right] \rho_{p,p'}.
\end{equation}
 It is easily seen that Eq. \eqref{energy_master_eqn} is a separable differential equation which can be solved to yield
%
%
%%%%%%%%%%%%%%%%%% equation %%%%%%%%%%%%%%%%%%%%%%%%%%%
\begin{equation}
\label{density_in_energy}
    \rho_{p,p'}(t) = {\rm exp}\left[ -\frac{i\left(E(p)  - E(p')\right)t}{\hbar}  - 2m\sigma \left( \Delta E^{3/2} \right)^2t\right]\rho_{p,p'}(0)
\end{equation}
from which one concludes the 
decoherence time to be 
%
%
%%%%%%%%%%%%%%%%%% equation %%%%%%%%%%%%%%%%%%%%%%%%%%%
\begin{equation}
\label{decoherence_time}
    \tau_D = \frac{1}{2m \sigma\left(\Delta E^{3/2}\right)^2} = \frac{\hbar^4}{8 m  t_p \ell_{p}^2 \left(\Delta E^{3/2}\right)^2}~.
\end{equation}
As expected, as long as the diagonal part, i.e., $\Delta E^{3/2}=0$, is concerned, there is no time evolution. However, when considering the off-diagonal terms (for which $\Delta E^{3/2}\neq0$), we can recognize an exponential damping that is a typical signature of the decoherence mechanism with a characteristic time which precisely equals the decoherence time.
In order to better emphasize the behavior of $\tau_D$, in Fig. \ref{fig:my_label1} we plot the decoherence time as a function of the energy and the mass of a given quantum system  to show that it gets smaller and smaller as the macroscopic scale is approached.
%
%
%
%%%%%%%%%%%%%%%%%% FIGURE %%%%%%%%%%%%%%%%%%%%%%%%%%%

\begin{figure}[ht]
    \centering
    \includegraphics[width=17cm, height=10cm]{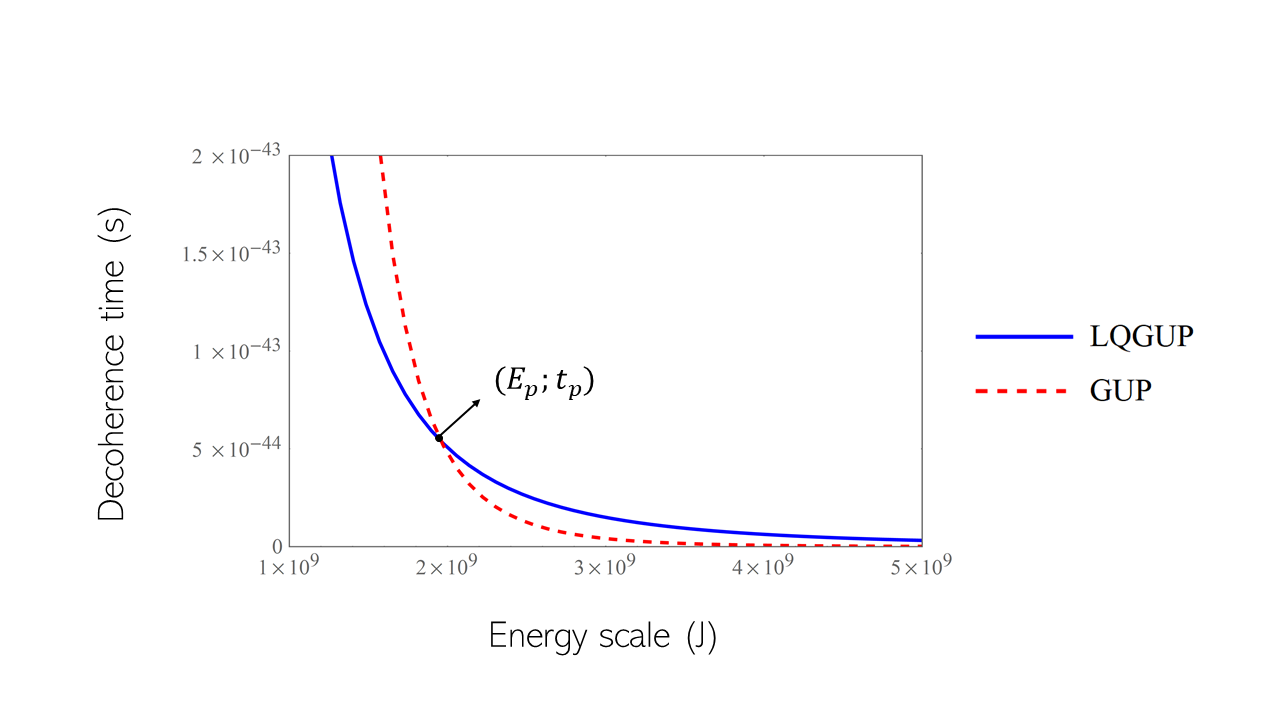}
    \caption{Decoherence time as a function of energy for the LQGUP (solid line) and for the GUP (dashed line).}
    \label{fig:my_label2}
\end{figure}

\noindent
It is noteworthy to mention that in Ref. \cite{Petruzziello:2020wkd}, the decoherence time obtained from considerations involving a GUP of the form shown in Eq. \eqref{QGUP} (which is only quadratic in the momentum) has the following shape
%
%
%%%%%%%%%%%%%%%%%% equation %%%%%%%%%%%%%%%%%%%%%%%%%%%
\begin{equation}
\label{Petruzziello_decoherence_time}
     \tau_D = \frac{\hbar^6}{16 m^2  t_p \ell_{p}^4 \left(\Delta E^{2}\right)^2}, \qquad \Delta E^{2} \equiv E^2(p)-E^2(p').
\end{equation}
In Fig. \ref{fig:my_label2}, we exhibit the comparison between the two forms of the decoherence times. As the picture clearly conveys, the decoherence time obtained from LQGUP arguments is shorter than the one obtained in Ref. \cite{Petruzziello:2020wkd} in the mesoscopic regime below the Planck size but it is longer beyond the same scale. To compare the two quantities, we have rephrased their expressions in terms of $\hbar$, $E_p$ and the average energy $E$ only, which gives
%%%%%%%%%%%%%%%%%% equation %%%%%%%%%%%%%%%%%%%%%%%%%%%
\begin{equation}\label{endif}
    \tau_{D, GUP} \simeq \frac{\hbar\, E_{p}^5}{E^6}, \qquad \tau_{D, LQGUP} \simeq \frac{\hbar\, E_{p}^2}{E^3}.
\end{equation}
%%%%%%%%%%%%%%%%%%%%%%%%%%%%%%%%%%%%%%%%%%%%%%%%%%%%%%%
%
%
%
%%%%%%%%%%%%%%%%%% FIGURE %%%%%%%%%%%%%%%%%%%%%%%%%%%
%
%
%%%%%%%%%%%%%%%%%%%%%%%%%%%%%%%%%%%%%%%%%%%%%%%%%%%%%%%
\section{Experimental Comparison}
%%%%%%%%%%%%%%%%%%%%%%%%%%%%%%%%%%%%%%%%%%%%%%%%%%%%%%
%
%
\par\noindent
We now extend the comparison between the two models of decoherence stemming from string theory and DSR-inspired GUP by thoroughly investigating their experimental implications. This will further corroborate the significant difference between the two predictions, which will appear to be macroscopic. To this aim, we employ the experimental setup based on cavity optomechanics already discussed in Refs. \cite{Petruzziello:2020wkd,Pfister:2015sna}. In a nutshell, the laboratory apparatus consists of two optomechanical cavities with two mirrors: in one of the cavities, both mirrors are fixed, whilst in the other only one mirror is fixed, the second being allowed to move so as to be subjected to gravitational decoherence. Atomic or molecular mechanical oscillators are trapped between the mirrors of both cavities by means of a harmonic potential and are prepared in an entangled state in order to act as correlated photon sources. Starting from this setup, the same external laser is then  switched on for both systems to excite the sources and to let them decay back to the ground state by emitting a photon in the process. The decoherence rate can thus be deduced by observing the vibration modes.

To theoretically understand the several steps of the decoherence phenomenon, the min-entropy  \cite{Pfister:2015sna}  can be exploited to provide a measure for the entanglement of the system with the environment. In our case, the two cavities $A$ and $B$, along with the environment $E$, can be outlined as tripartite system. %A measure of decoherence consists of a measurement of parameter $\beta$ which sets an upper limit on the decoherence rate of system A to the environment \cite{Pfister:2015sna}
%%%%%%%%%%%%%%%%%% equation %%%%%%%%%%%%%%%%%%%%%%%%%%%
%\bea
%\text{Dec}(A|E) \leq h(\beta)
%\eea
%%%%%%%%%%%%%%%%%%%%%%%%%%%%%%%%%%%%%%%%%%%%%%%%%%%%%%%
% describes the relationship between min-entropy and decoherence rate as follows:
By preparing the systems $A$ and $B$ in a maximally entangled state, one can account for the decoherence of the system $A$ by witnessing the loss of entanglement in the bipartite system $AB$. A mathematical function that quantifies the above process is defined as \cite{Pfister:2015sna}  
%%%%%%%%%%%%%%%%%% equation %%%%%%%%%%%%%%%%%%%%%%%%%%%
\bea\label{deae}
\text{Dec}(A|E) = \max \limits_{R_{E\rarrow B}} F^2(\Phi_{A,B},\mathbb{1}_{A}\otimes R_{E\rarrow B}(\rho_{AE})),
\eea
%%%%%%%%%%%%%%%%%%%%%%%%%%%%%%%%%%%%%%%%%%%%%%%%%%%%%%%
where $\Phi_{A,B}$ is a maximally entangled state $AB$, $R_{E\rarrow B}$ is the set of all quantum operations from $E$ to $B$, $\rho_{AE}$ is the reduced state describing the bipartite subsystem $AE$ and $F$ is the Uhlmann fidelity given by
%%%%%%%%%%%%%%%%%% equation %%%%%%%%%%%%%%%%%%%%%%%%%%%
\bea
F(\rho,\sigma) = \text{Tr}\sqrt{\sqrt{\rho}~\sigma \sqrt{\rho}}.
\eea
%%%%%%%%%%%%%%%%%%%%%%%%%%%%%%%%%%%%%%%%%%%%%%%%%%%%%%%
In the ideal case in which there is no interaction with the environment, no decoherence occurs; this entails that $A$ and $B$ remain maximally entangled and no correlation is established between $A$ and $E$. On the other hand, a more realistic picture contemplates the existence of a decoherence mechanism, which thus can be interpreted as a degradation of the perfect correlation between $A$ and $B$ and a partial correlation between $A$ and $E$. As already anticipated, the min-entropy represents a faithful measure for the above process since it is associated to the decoherence rate of Eq.\eqref{deae} via the relation \cite{Pfister:2015sna}
%%%%%%%%%%%%%%%%%% equation %%%%%%%%%%%%%%%%%%%%%%%%%%%
\bea
H_{\text{min}}(A|E)  = -\log_2\left[{d_{A}\text{Dec}(A|E)}\right],
\eea
%%%%%%%%%%%%%%%%%%%%%%%%%%%%%%%%%%%%%%%%%%%%%%%%%%%%%%%
with $d_A$ being the dimension of the Hilbert space of $A$.
In light of the previous reasoning, we can conclude that the entanglement between the systems $A$ and $B$ can be regarded as the observable quantity to measure to overcome the challenge of detecting the decoherence arising from the interactions between one of the subsystems and the environment. 

Now, in order to quantify the amount of correlations shared by parties $A$ and $B$, we make use of the standard Clauser-Horne-Shimony-Holt (CHSH) nonlocal correlation measurement\footnote{We stress that, in principle, we could also rely on other forms of Bell's inequality to achieve the same outcome \cite{Pfister:2015sna}.}, which allows us to compare our results with the ones of Ref. \cite{Petruzziello:2020wkd}. In a simple configuration, both systems $A$ and $B$ are measured using a set of two observables, $A_0$ and $A_1$ for the subsystem $A$ and $B_0$ and $B_1$ for the subsystem $B$. According to this scenario, the correlation parameter then reads
%%%%%%%%%%%%%%%%%% equation %%%%%%%%%%%%%%%%%%%%%%%%%%%
\bea
\xi = tr[ \rho_{AE}(A_0 \otimes B_0 + A_0 \otimes B_1 + A_1 \otimes B_0 -A_1 \otimes B_1)],
\eea
%%%%%%%%%%%%%%%%%%%%%%%%%%%%%%%%%%%%%%%%%%%%%%%%%%%%%%%
where we have $A_0 = \sigma_x$, $A_1 = \sigma_z$, $B_0=(\sigma_{x}-\sigma_z)/{\sqrt{2}}$, and $B_1=(\sigma_{x}+\sigma_z)/{\sqrt{2}}$, with $\sigma_i$ being the $i-$th Pauli matrix.
In the aforementioned setting, two phenomena are expected to simultaneously occur: gravitational decoherence and mechanical heating. The latter effect is undesirable (but unavoidable in a realistic setting), since it interferes with the direct observation of gravitational decoherence and needs to be suppressed as much as possible to properly test our model. Now, it can be shown \cite{Pfister:2015sna} that the gravitational decoherence and mechanical heating rates are respectively given by
%%%%%%%%%%%%%%%%%% equation %%%%%%%%%%%%%%%%%%%%%%%%%%%
\begin{equation}
    \Lambda_{grav} = \frac{1}{\tau_{d}} = \frac{8 m  t_p \ell_{p}^2 \left(\Delta E^{3/2}\right)^2}{\hbar^4}, \qquad \Lambda_{heat} = \frac{k_{B}T}{\hbar Q},
\end{equation}
%%%%%%%%%%%%%%%%%%%%%%%%%%%%%%%%%%%%%%%%%%%%%%%%%%%%%%%
with $k_B$ being the Boltzmann constant, $T$ the temperature and $Q$ the quality factor of the cavity. Using these definitions, the decoherence rate for the entangled system can thus be derived (see Ref. \cite{Pfister:2015sna} for more details) and it is given by
%%%%%%%%%%%%%%%%%% equation %%%%%%%%%%%%%%%%%%%%%%%%%%%
\begin{equation}
    \text{Dec}(A|E) = \frac{1}{4}\left(1+\sqrt{1-\exp\left(-4(1+2\bar{n})\frac{g_{0}^2}{\omega_{m}^2}\sin^2\left(\frac{\omega_{m}t}{2}\right) \right)} \right),
\end{equation}
%%%%%%%%%%%%%%%%%%%%%%%%%%%%%%%%%%%%%%%%%%%%%%%%%%%%%%%
where 
%%%%%%%%%%%%%%%%%% equation %%%%%%%%%%%%%%%%%%%%%%%%%%%
\begin{equation}
    \bar{n} = \frac{2 \Lambda_{grav}}{\gamma_{m}} + \frac{2\Lambda_{heat}}{\gamma_{m}}
\end{equation}
%%%%%%%%%%%%%%%%%%%%%%%%%%%%%%%%%%%%%%%%%%%%%%%%%%%%%%%
is the average phonon number, $g_{0}$ is the single-photon optomechanical coupling rate (typically $g_{0}\sim 1 s^{-1}$), $\omega_m$ is the frequency of the trapping harmonic potential and $\gamma_m=\omega_m/Q$. Given a quality factor of $
 Q\sim 10^{10}$ for the cavity, the following values represent a reasonable choice for the experimental parameters to measure the gravitational decoherence: $\omega_{m} \sim 1$ s$^{-1}$, $\gamma_{m} \sim 10^{-10}$ s$^{-1}$, T $\sim$ 1 nK and $E\sim10$ J, which corresponds to the rest energy of a molecular structure with a mass of the order of $10^{-16}$ Kg. Such a choice for the mass coincides with the one required for the implementation of devised laboratory tests whose aim is to witness the quantum nature of the gravitational interaction \cite{bose,vedral}.
 
\begin{figure}[ht]
    \centering
    \includegraphics[width=18cm, height=10cm]{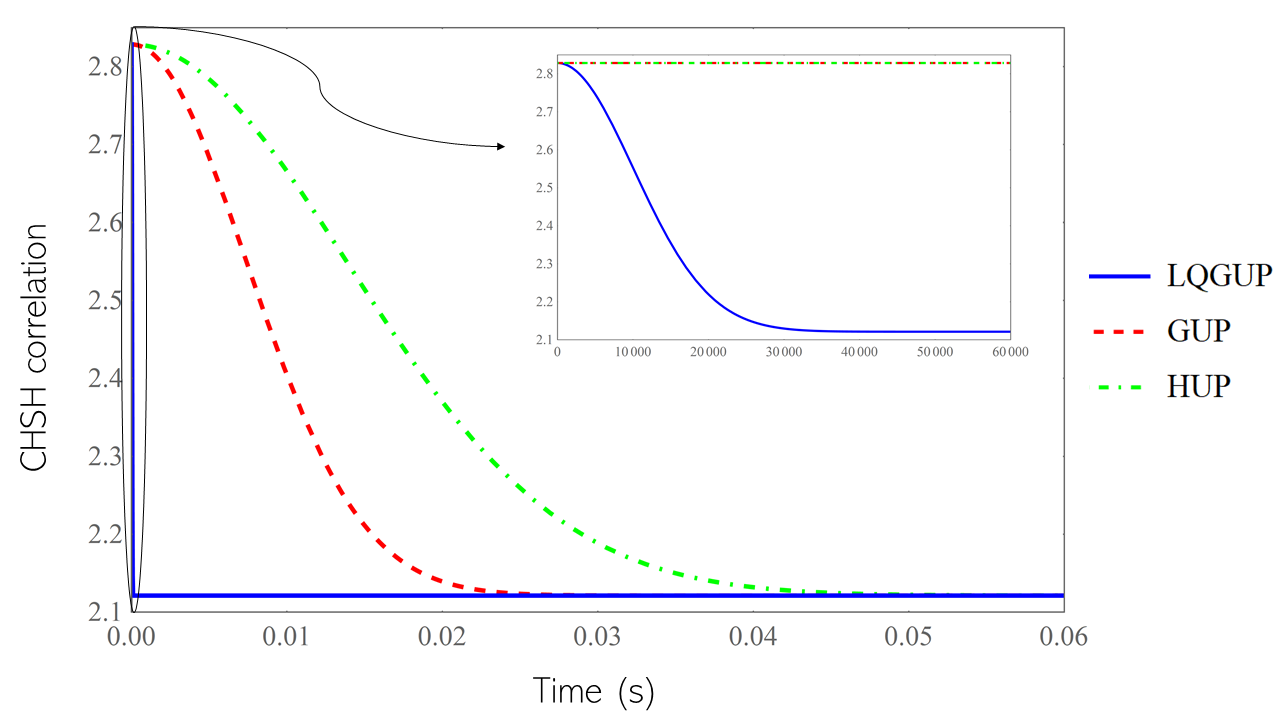}
    \caption{Behavior of the CHSH correlation as a function of time for the LQGUP (solid line), the quadratic GUP (dashed line) and the unperturbed case (dot-dashed line). The experimental values used to generate the plot are the ones introduced in the main text of the paper. In the inset, the curve associated with the LQGUP model has been magnified by a factor of $10^{37}$ to let it be visually compatible with the other scenarios.}
    \label{fig2}
\end{figure}

Bearing this in mind, we are finally able to plot the CHSH correlation parameter for our LQGUP model and compare it with the result stemming from the quadratic GUP contained in Ref. \cite{Petruzziello:2020wkd} as well as the unperturbed case in which there is only mechanical heating. The outcome of such a comparison is summarized in Fig. \ref{fig2}. 
%%%%%%%%%%%%%%%%%% FIGURE %%%%%%%%%%%%%%%%%%%%%%%%%%%
%%%%%%%%%%%%%%%%%%%%%%%%%%%%%%%%%%%%%%%%%%%%%%%%%%%%%
The plot shows a  significant behavioral discrepancy between the two models. Indeed, the effect of considering the decoherence due to quadratic GUP amounts to a small (but still detectable) reduction of the time it takes for the systems $A$ and $B$ to lose their correlation in comparison with the unperturbed case. On the other hand, the degradation of the same correlation in the framework of LQGUP is almost instantaneous. The reason for this difference lies in the distinct energy dependence of the decoherence times associated with the two GUP models, which has already been explored in Fig. \ref{fig:my_label2} and Eq. \eqref{endif}. However, in terms of experimentally observed quantities, the gap between the two predictions is so sharp that the two scenarios become clearly distinguishable with a laboratory test.

%This achievement opens up a wide spectrum of possibilities to gain an insight on the quantum gravitational models behind the analyzed GUPs, as their predictions can be feasibly investigated in an implementable laboratory test. 
On a final note, regarding the LQGUP implications studied in the present paper, it must be said that, due to the energy dependence of its decoherence time, the molecular sources which can be used in experiments do not have to be necessarily big to give rise to detectable outcomes. As a matter of fact, whilst in the quadratic GUP case we would need masses of the order of $10^{-16}$ Kg, i.e., the reference value with which Fig. \ref{fig2} has been drawn, for the current case we can reach experimentally accessible timescales with sources having a mass of $10^{-25}$ Kg. With a similar setting, the curve associated with the LQGUP decoherence of Fig. \ref{fig2} would take $\mathcal{O}(10^{-2})$ s to be reduced by a factor $1/e$, thus allowing for its experimental validation or falsification even with small molecules (which of course are much easier to handle with respect to bigger ones).
%%%%%%%%%%%%%%%%%%%%%%%%%%%%%%%%%%%%%%%%%%%%%%%%%%%%%%%
\section{Discussion}
%%%%%%%%%%%%%%%%%%%%%%%%%%%%%%%%%%%%%%%%%%%%%%%%%%%%%%
%
%
%
\par\noindent
In this paper, we have shown that decoherence phenomena arising from quantum gravity models compatible with different forms of generalized uncertainty principles can be efficiently understood and compared by deriving the respective master evolution equations and decoherence times and by tracking the corresponding decoherent dynamics of nonlocal quantum correlations of the CHSH type.
We have studied the two most relevant versions of the GUP arising from string theory and doubly special relativity, respectively. We have exhibited that, in addition to the quadratic GUP already investigated in Ref. \cite{Petruzziello:2020wkd}, also the GUP with linear and quadratic terms in the momentum operator 
yields a viable explanation for the quantum-to-classical transition. 

This mechanism can be realized by assuming a stochastic nature for the deformation parameter, encoding the foamy structure of spacetime at the Planck scale. This very consideration allows to derive a Lindblad-type master equation,
from which one can deduce several physical consequences of the GUP-induced gravitational decoherence, such as the entropy variation and decoherence time. Specifically, we have seen how the linear entropy monotonically increases with time, thus leading to a continuous increase of the state mixedness and to an irreversible evolution process. On the other hand, the decoherence time turns out to decrease for increasing energy and mass of the system. This occurs at a rate faster than the decoherence due to GUP without linear momentum term (as derived in Ref. \cite{Petruzziello:2020wkd}) for mesoscopic regimes beyond the Planck scale, whilst the same trend is inverted below it. Such features suggest that there exists a strong relation between a particular structure of the GUP and the ensuing decoherence time.

The dependence of the gravitational decoherence on the actual form of the GUP becomes even more cogent when looking at the decoherent time evolution of nonlocal quantum correlations of the CHSH type entering the Bell inequalities.
As a matter of fact, we have seen that, according to the employed GUP version, for an accurately tuned laboratory test involving cavity optomechanics we derive completely different outcomes. Most importantly, we have shown how it is possible to achieve a detectable deviation between the GUP and LQGUP predicted dynamics of the CHSH nonlocal correlations on time scales that are experimentally accessible. Indeed, by selecting molecular structures with masses of the order of $10^{-16}$ Kg, it is easy to check that the temporal discrepancy between the two different GUP schemes is about a hundredth of a second (as outlined in Fig. \ref{fig2}). Assuming that the above large molecules could be experimentally prepared in a superposition as envisaged in Refs. \cite{bose,vedral}, one could potentially be able to falsify one GUP version in favor of the other. Such a finding would in turn have a great impact on the underlying quantum gravitational theories, i.e., string theory and DSR, thereby shedding light on the possibility of discriminating quantum gravity models by means of non-relativistic quantum tests. Should the preparation of massive molecular structures be still difficult in the forthcoming future, at least for the LQGUP framework we can rely on smaller sources, as its effects in terms of decoherence would be appreciable even when the mass range is of the order of $10^{-25}$ Kg, thus opening up promising experimental opportunities.

In conclusion, depending on the agreement with the experimental data, not only one can determine which model is the most suited to consistently describe the quantum-to-classical transition, but most importantly one can establish a tool for testing and falsifying different theoretical approaches. Furthermore, because of the impossibility of experimentally reaching high-energy regimes to probe quantum gravitational effects directly, the aforementioned non-relativistic laboratory test could be deemed as a promising candidate to verify the quantumness of gravity.

Ultimately, the framework introduced in the present work can be applied to identify and select classes of phenomenologically meaningful uncertainty relations, capable of explaining the quantum-to-classical transition with quantum gravitational effects. Since the decoherence mechanism stemming from the GUP strictly depends on the deviation from the HUP, future high-precision laboratory probes may be able to provide valuable evidence, essential for the construction of consistent and experimentally testable models of quantum gravity. Even if such experiments are not immediately realized, one would potentially enjoy strong bounds on the large number of generalized uncertainty principles currently available, that are still inaccessible to direct experimental verification and whose theoretical background is still a matter of intense debate.
%
%
%
%%%%%%%%%%%%%%%%%%%%%%%%%%%%%%%%%%%%%%%%%%%%%%%%%%%%%%
%
%
%%%%%%%%%%%%%%%%%%%%%%%%%%%%%%%%%%%%%%%%%%%%%%%%%%%%%%
\section{Acknowledgement}
%%%%%%%%%%%%%%%%%%%%%%%%%%%%%%%%%%%%%%%%%%%%%%%%%%%%%%
%
\par\noindent
This work was supported by the Natural Sciences and Engineering Research Council of Canada.
F.I. and L.P. acknowledge support by MUR (Ministero dell'Universit\`a e della Ricerca) via the project PRIN 2017 ``Taming complexity via QUantum Strategies: a Hybrid Integrated Photonic approach'' (QUSHIP) Id. 2017SRNBRK.
S.D., E.C.V. and L.P. would like to acknowledge networking support by the COST Action CA18108. L.P.  is thankful to the ``Angelo Della Riccia'' foundation for the awarded fellowship received to support the study at Universit\"at Ulm.
%
%%%%%%%%%%%%%%%%%%%%%%%%%%%%%%%%%%%%%%%%%%%%%%%%%%%%%%
%
\appendix
%
%
%%%%%%%%%%%%%%%%%%%%%%%%%%%%%%%%%%%%%%%%%%%%%%%%%%%%%%
\section{Second-order perturbation of \texorpdfstring{$\alpha$}{Lg} in the Hamiltonian}
%%%%%%%%%%%%%%%%%%%%%%%%%%%%%%%%%%%%%%%%%%%%%%%%%%%%%%
\label{appendix_A}
\par\noindent
In this Appendix, we show how the main outcome of the current work is left untouched even if we push our analysis beyond the first-order approximation.
Indeed, if we consider second-order approximations in $\alpha$ for $H_{1}$ in the Hamiltonian of Eq. \eqref{full_LQGUP}, we get
%
%
%%%%%%%%%%%%%%%%%% equation %%%%%%%%%%%%%%%%%%%%%%%%%%%
\begin{equation}
\label{Appendix_Eqn:Schrodinger_eqn}
    i \hbar \partial_t |\psi\rangle = H|\psi\rangle = \left(H_0 + H_{1} + \mathcal{O}(\alpha^3) \right) |\psi\rangle,
\end{equation}
with
%
%
%%%%%%%%%%%%%%%%%% equation %%%%%%%%%%%%%%%%%%%%%%%%%%%
\begin{equation}
\label{Appendix_Eqn:Hamiltonian_definitions}
H_0 = \frac{p^2_0}{2m} + V, \ \  H_{1} = -\frac{\alpha}{m} p_{0}^3 + \frac{5\alpha^{2}}{2m}p_{0}^4~,
\end{equation}
where we ignore terms that go like $\mathcal{O}(\alpha^3)$.
\par\noindent
We note that the addition of the second order $\alpha$ term will begin to have a significant effect only in correspondence of Eq. \eqref{average_liouvile_von_neumann}, which can be written as
%
%
%%%%%%%%%%%%%%%%%% equation %%%%%%%%%%%%%%%%%%%%%%%%%%%
\begin{equation}
\label{Appendix_Eqn:long_Liouville}
\partial_t \tilde{\rho}(t) =   - \frac{1}{\hbar^2} \int_{0}^{t} \left \langle \tilde{H}_{1}(t)\tilde{H}_{1}(t')\tilde{\varrho}(t) -\tilde{H}_{1}(t)\tilde{\varrho}(t)\tilde{H}_{1}(t') - \tilde{H}_{1}(t')\tilde{\varrho}(t)\tilde{H}_{1}(t) + \tilde{\varrho}(t)\tilde{H}_{1}(t')\tilde{H}_{1}(t) \right \rangle dt'.
\end{equation}
In this case, if we substitute the new value for $H_{1}$ we obtain
%
%
%%%%%%%%%%%%%%%%%% equation %%%%%%%%%%%%%%%%%%%%%%%%%%%
\bea
\label{Appendix_eqn:The_larget_equation}
\partial_t \tilde{\rho}(t) &=&   - \frac{1}{\hbar^2} \int_{0}^{t}  \bigg \langle \left(\frac{\alpha(t)}{m} \tilde{p}_{0}^3(t) + \frac{5\alpha^{2}(t)}{2m}\tilde{p}_{0}^4(t) \right) \left(\frac{\alpha(t')}{m} \tilde{p}_{0}^3(t') + \frac{5\alpha^{2}(t')}{2m}\tilde{p}_{0}^4(t') \right)\tilde{\varrho}(t)\nonumber\\
&&- \left(\frac{\alpha(t)}{m} \tilde{p}_{0}^3(t) + \frac{5\alpha^{2}(t)}{2m}\tilde{p}_{0}^4(t) \right) \tilde{\varrho}(t)\left(\frac{\alpha(t')}{m} \tilde{p}_{0}^3(t')+\frac{5\alpha^{2}(t')}{2m}\tilde{p}_{0}^4(t') \right)\nonumber\\ 
&&- \left(\frac{\alpha(t')}{m} \tilde{p}_{0}^3(t') + \frac{5\alpha^{2}(t')}{2m}\tilde{p}_{0}^4(t') \right) \tilde{\varrho}(t) \left(\frac{\alpha(t)}{m} \tilde{p}_{0}^3(t) + \frac{5\alpha^{2}(t)}{2m}\tilde{p}_{0}^4(t) \right)\nonumber\\
&&+  \tilde{\varrho}(t) \left(\frac{\alpha(t')}{m} \tilde{p}_{0}^3(t') + \frac{5\alpha^{2}(t')}{2m}\tilde{p}_{0}^4(t') \right) \left(\frac{\alpha(t)}{m} \tilde{p}_{0}^3(t) + \frac{5\alpha^{2}(t)}{2m}\tilde{p}_{0}^4(t) \right) \bigg \rangle dt'.
\eea 
If we keep the terms up to $\mathcal{O}(\alpha^{2})$, we are left with
%
%
%%%%%%%%%%%%%%%%%% equation %%%%%%%%%%%%%%%%%%%%%%%%%%%
\begin{equation}
\label{Appendix_eqn:reduced_long_equation}
\partial_t \tilde{\rho}(t) =   - \frac{4}{\hbar^2} \int_{0}^{t} \left \langle \alpha(t)\alpha(t')\left(\frac{\tilde{p}^{3}_{0}(t)}{2m}\frac{\tilde{p}^{3}_{0}(t')}{2m}\tilde{\varrho}(t)  - \frac{\tilde{p}^{3}_{0}(t)}{2m} \tilde{\varrho}(t)\frac{\tilde{p}^{3}_{0}(t')}{2m} -\frac{\tilde{p}^{3}_{0}(t')}{2m} \tilde{\varrho}(t) \frac{\tilde{p}^{3}_{0}(t)}{2m} +  \tilde{\varrho}(t)\frac{\tilde{p}^{3}_{0}(t')}{2m}\frac{\tilde{p}^{3}_{0}(t)}{2m}\right) \right \rangle dt'
\end{equation}
%
%
%%%%%%%%%%%%%%%%%% equation %%%%%%%%%%%%%%%%%%%%%%%%%%%
\begin{equation}
\label{Appendix_eqn:Liouville_in_commutators}
    \partial_t \tilde{\rho}(t) =   - \frac{4}{\hbar^2} \int_{0}^{t}\langle\alpha(t)\alpha(t')\rangle \left [\frac{\tilde{p}^{3}_{0}(t)}{2m}, [\frac{\tilde{p}^{3}_{0}(t')}{2m},\tilde{\rho}(t) ]\right ] dt' 
\end{equation}
that is the same integral obtained in Eq. \eqref{prefinal_integral_version} in which we only considered terms up to $\mathcal{O}(\alpha)$. Therefore, by following the same steps and recalling that $\alpha(t) = {\alpha_0 \ell_{p}}/{\hbar}={\ell_{p}\sqrt{t_{p}}\chi(t)}/{\hbar}$, we obtain
%
%
%%%%%%%%%%%%%%%%%% equation %%%%%%%%%%%%%%%%%%%%%%%%%%%
\begin{equation}
\label{Appendix_eqn:liouville_fluctuations_inserted}
\partial_t \tilde{\rho}(t) = - \sigma \int_{0}^{t}\left \langle \chi_0(t) \chi_0(t') \right \rangle \left[\frac{\tilde{p}^3_0(t)}{2m},\left[\frac{\tilde{p}^3_0(t')}{2m},\tilde{\rho}(t)\right]\right]  dt',
\end{equation}
which becomes
%
%
%%%%%%%%%%%%%%%%%% equation %%%%%%%%%%%%%%%%%%%%%%%%%%%
\begin{equation}
\label{Appendix_eqn:liouville_delta_function}
\partial_t \tilde{\rho}(t) = - \sigma \int_{0}^{t} \delta(t-t') \left[\frac{\tilde{p}^3_0(t)}{2m},\left[\frac{\tilde{p}^3_0(t')}{2m},\tilde{\rho}(t)\right]\right]  dt'~.
\end{equation}
Finally, we derive the expression
%
%
%%%%%%%%%%%%%%%%%% equation %%%%%%%%%%%%%%%%%%%%%%%%%%%
\begin{equation}
\label{Appendix_eqn:final_liouville}
 \partial_t \tilde{\rho}(t) =   -\sigma \left[\frac{\tilde{p}^3_0(t)}{2m}, \left[\frac{\tilde{p}^3_0(t)}{2m},\tilde{\rho}(t) \right]\right]
\end{equation}
which is precisely Eq. \eqref{master_eqn_momentum_representation}.
\par\noindent
By recovering the Schr\"odinger representation and going back to the Liouville-von Neumann equation, we can retrieve the Lindblad-type master equation, i.e., Eq. \eqref{lindbladian},  by vanishing the external potential
%
%
%%%%%%%%%%%%%%%%%% equation %%%%%%%%%%%%%%%%%%%%%%%%%%%
\begin{equation}
\label{Appendix_eqn:master_eqn}
\partial_t \rho(t) = -\frac{i}{\hbar} \left[\frac{p_0^2}{2m},\rho(t)\right]  - \sigma  \left[\frac{\tilde{p}^3_0}{2m},\left[\frac{\tilde{p}^3_0}{2m},\rho(t)\right]\right].
\end{equation}
Because of the appearance of a double Hamiltonian in Eq. \eqref{Appendix_Eqn:long_Liouville}, it is clear that we can safely neglect all higher-order terms above $\mathcal{O}(\alpha)$, which by the way are extremely suppressed due to the smallness of the corrections compared to the unmodified scenario.
%
%
%
%
%
%
%
%%%%%%%%%%%%%%%%%%%%%%%%%%%%%%%%%%%%%%%%%%%%%%%%%%%

%%%%%%%%%%%%%%%%%%%%%%%%%%%%%%%%%%%%%%%%%%%%%%%%%%%
%
%
%
%
%%%%%%%%%%%%%%%%%%%%%%%%%%%%%%%%%%%%%%%%%%%%%%%%%%%%

\begin{thebibliography}{99}

%\cite{Zeh:1970zz}
\bibitem{Zeh:1970zz}
H.~D.~Zeh,
%``On the interpretation of measurement in quantum theory,''
Found. Phys. \textbf{1}, 69-76 (1970).
%doi:10.1007/BF00708656
%236 citations counted in INSPIRE as of 18 Aug 2021


%\cite{Zurek:2003zz}
\bibitem{Zurek:2003zz}
W.~H.~Zurek,
%``Decoherence, einselection, and the quantum origins of the classical,''
Rev. Mod. Phys. \textbf{75}, 715-775 (2003)
%doi:10.1103/RevModPhys.75.715
[arXiv:quant-ph/0105127 [quant-ph]].
%436 citations counted in INSPIRE as of 18 Aug 2021


%\cite{Anastopoulos:2000hg}
\bibitem{Anastopoulos:2000hg}
C.~Anastopoulos,
%``Frequently asked questions about decoherence,''
Int. J. Theor. Phys. \textbf{41}, 1573-1590 (2002)
%doi:10.1023/A:1020144800650
[arXiv:quant-ph/0011123 [quant-ph]].
%9 citations counted in INSPIRE as of 18 Aug 2021


%\cite{Hornberger}
\bibitem{Hornberger}
K.~Hornberger,
%``Introduction to Decoherence Theory,''
%doi:10.1007/978-3-540-88169-8_5
Lect. Notes Phys. \textbf{768}, 221-276 (2009)
[arXiv:quant-ph/0612118 [quant-ph]].
%2 citations counted in INSPIRE as of 18 Aug 2021


%\cite{Schlosshauer:2019ewh}
\bibitem{Schlosshauer:2019ewh}
M.~Schlosshauer,
%``Quantum Decoherence,''
Phys. Rept. \textbf{831}, 1-57 (2019)
%doi:10.1016/j.physrep.2019.10.001
[arXiv:1911.06282 [quant-ph]].
%16 citations counted in INSPIRE as of 18 Aug 2021


%\cite{Schlosshauer:2003zy}
\bibitem{Schlosshauer:2003zy}
M.~Schlosshauer,
%``Decoherence, the Measurement Problem, and Interpretations of Quantum Mechanics,''
Rev. Mod. Phys. \textbf{76}, 1267-1305 (2004)
%doi:10.1103/RevModPhys.76.1267
[arXiv:quant-ph/0312059 [quant-ph]].
%232 citations counted in INSPIRE as of 18 Aug 2021


%\cite{Kempf:1994su}
\bibitem{Kempf:1994su}
A.~Kempf, G.~Mangano and R.~B.~Mann,
%``Hilbert space representation of the minimal length uncertainty relation,''
Phys. Rev. D \textbf{52}, 1108-1118 (1995)
%doi:10.1103/PhysRevD.52.1108
[arXiv:hep-th/9412167 [hep-th]].
%1313 citations counted in INSPIRE as of 18 Aug 2021

\bibitem{noncom2} 
T.~Kanazawa, G.~Lambiase, G.~Vilasi, and A.~Yoshioka, Eur. Phys. J. C \textbf{79}, 95 (2019).

\bibitem{noncom3}
G.G.~Luciano and L.~Petruzziello,
  %``GUP parameter from Maximal Acceleration,''
  Eur.\ Phys.\ J.\ C {\bf 79}, 283 (2019) [arXiv:1902.07059 [hep-th]].

\bibitem{bis}
R.J.~Adler, P.~Chen, and D.I.~Santiago, Gen. Rel. Grav. \textbf{33}, 2101 (2001) [arXiv:gr-qc/0106080 [gr-qc]].

\bibitem{quar}
P.~Chen, Y.C.~Ong, and Dh.~Yeom, Phys. Rep. \textbf{603}, 1 (2015) [arXiv:1412.8366 [gr-qc]].

\bibitem{ses}
F.~Scardigli, G.~Lambiase, and E.~Vagenas, Phys. Lett. B \textbf{767}, 242 (2017) [arXiv:1611.01469 [hep-th]].

\bibitem{ong}
P.~Chen, Y.~C.~Ong and D.~h.~Yeom,
%``Generalized Uncertainty Principle: Implications for Black Hole Complementarity,''
JHEP \textbf{12}, 021 (2014) [arXiv:1408.3763 [hep-th]].

\bibitem{set}
L.~Buoninfante, G.G.~Luciano, and L.~Petruzziello,
  %``Generalized Uncertainty Principle and Corpuscular Gravity,''
  Eur.\ Phys.\ J.\ C {\bf 79}, 663 (2019) [arXiv:1903.01382 [gr-qc]].
  
\bibitem{ot}
  L.~Buoninfante, G.~Lambiase, G.G.~Luciano, and L.~Petruzziello,
  %``Phenomenology of GUP stars,''
  Eur. Phys. J. C \textbf{80}, 853 (2020) [arXiv:2001.05825 [gr-qc]].
	

\bibitem{dieci}
L.~Petruzziello,
%``Generalized uncertainty principle with maximal observable momentum and no minimal length indeterminacy,''
Class. Quant. Grav. \textbf{38}, 135005 (2021) [arXiv:2010.05896 [hep-th]].


\bibitem{qft}
S.~Hossenfelder,
  %``Minimal Length Scale Scenarios for Quantum Gravity,''
  Living Rev.\ Rel.\  {\bf 16}, 2 (2013) [arXiv:1203.6191 [gr-qc]].

\bibitem{qft4}
M.~Blasone, G.~Lambiase, G.G.~Luciano, L.~Petruzziello, and F.~Scardigli,	
  Int.\ J.\ Mod.\ Phys.\ D {\bf 29}, 2050011 (2020) [arXiv:1912.00241 [hep-th]].	

\bibitem{plenio2}
S.P.~Kumar and M.B.~Plenio,
%``On Quantum Gravity Tests with Composite Particles,''
Nature Commun. \textbf{11}, 3900 (2020) [arXiv:1908.11164 [quant-ph]].

%\cite{Bosso:2022vlz}
\bibitem{ourplb}
P.~Bosso, L.~Petruzziello and F.~Wagner,
%``The minimal length is physical,''
Phys. Lett. B \textbf{834}, 137415 (2022)
[arXiv:2206.05064 [gr-qc]].
%12 citations counted in INSPIRE as of 10 May 2023

%\cite{Jizba:2022icu}
\bibitem{ourprd}
P.~Jizba, G.~Lambiase, G.~G.~Luciano and L.~Petruzziello,
%``Decoherence limit of quantum systems obeying generalized uncertainty principle: New paradigm for Tsallis thermostatistics,''
Phys. Rev. D \textbf{105}, no.12, L121501 (2022)
[arXiv:2201.07919 [hep-th]].
%15 citations counted in INSPIRE as of 10 May 2023

%\cite{Das:2009hs}
\bibitem{Das:2009hs}
S.~Das and E.~C.~Vagenas,
%``Phenomenological Implications of the Generalized Uncertainty Principle,''
Can. J. Phys. \textbf{87}, 233-240 (2009)
%doi:10.1139/P08-105
[arXiv:0901.1768 [hep-th]].
%189 citations counted in INSPIRE as of 18 Aug 2021


%\cite{Das:2010sj}
\bibitem{Das:2010sj}
S.~Das and E.~C.~Vagenas,
%``Reply to `Comment on `Universality of Quantum Gravity Corrections' ',''
Phys. Rev. Lett. \textbf{104}, 119002 (2010)
%doi:10.1103/PhysRevLett.104.119002
[arXiv:1003.3208 [hep-th]].
%25 citations counted in INSPIRE as of 18 Aug 2021

%\cite{Adler:1999bu}
\bibitem{rnd}
R.~J.~Adler and D.~I.~Santiago,
%``On gravity and the uncertainty principle,''
Mod. Phys. Lett. A \textbf{14}, 1371 (1999)
doi:10.1142/S0217732399001462
[arXiv:gr-qc/9904026 [gr-qc]].
%407 citations counted in INSPIRE as of 10 May 2023

%\cite{Casadio:2022opg}
\bibitem{rnd2}
R.~Casadio, W.~Feng, I.~Kuntz and F.~Scardigli,
%``Minimum length (scale) in quantum field theory, generalized uncertainty principle and the non-renormalisability of gravity,''
Phys. Lett. B \textbf{838}, 137722 (2023)
doi:10.1016/j.physletb.2023.137722
[arXiv:2210.12801 [hep-th]].
%2 citations counted in INSPIRE as of 10 May 2023

%\cite{Amati:1987wq}
\bibitem{gener}
D.~Amati, M.~Ciafaloni and G.~Veneziano,
%``Superstring Collisions at Planckian Energies,''
Phys. Lett. B \textbf{197}, 81 (1987)
%809 citations counted in INSPIRE as of 10 May 2023

%\cite{Maggiore:1993rv}
\bibitem{gener2}
M.~Maggiore,
%``A Generalized uncertainty principle in quantum gravity,''
Phys. Lett. B \textbf{304}, 65-69 (1993)
[arXiv:hep-th/9301067 [hep-th]].
%882 citations counted in INSPIRE as of 10 May 2023

%\cite{Scardigli:1999jh}
\bibitem{gener3}
F.~Scardigli,
%``Generalized uncertainty principle in quantum gravity from micro - black hole Gedanken experiment,''
Phys. Lett. B \textbf{452}, 39-44 (1999)
[arXiv:hep-th/9904025 [hep-th]].
%704 citations counted in INSPIRE as of 10 May 2023

%\cite{Allali:2021puy}
\bibitem{Allali:2021puy}
I.~J.~Allali and M.~P.~Hertzberg,
%``General Relativistic Decoherence with Applications to Dark Matter Detection,''
Phys. Rev. Lett. \textbf{127}, no.3, 031301 (2021)
%doi:10.1103/PhysRevLett.127.031301
[arXiv:2103.15892 [gr-qc]].
%2 citations counted in INSPIRE as of 18 Aug 2021


%\cite{Anastopoulos:2013zya}
\bibitem{Anastopoulos:2013zya}
C.~Anastopoulos and B.~L.~Hu,
%``A Master Equation for Gravitational Decoherence: Probing the Textures of Spacetime,''
Class. Quant. Grav. \textbf{30}, 165007 (2013)
%doi:10.1088/0264-9381/30/16/165007
[arXiv:1305.5231 [gr-qc]].
%53 citations counted in INSPIRE as of 18 Aug 2021


%\cite{Asprea:2019dok}
\bibitem{Asprea:2019dok}
L.~Asprea, G.~Gasbarri and A.~Bassi,
%``Gravitational decoherence: A general nonrelativistic model,''
Phys. Rev. D \textbf{103}, no.10, 104041 (2021)
%doi:10.1103/PhysRevD.103.104041
[arXiv:1905.01121 [quant-ph]].
%10 citations counted in INSPIRE as of 18 Aug 2021


%\cite{Bonder:2015hja}
\bibitem{Bonder:2015hja}
Y.~Bonder, E.~Okon and D.~Sudarsky,
%``Can gravity account for the emergence of classicality?,''
Phys. Rev. D \textbf{92}, no.12, 124050 (2015)
%doi:10.1103/PhysRevD.92.124050
[arXiv:1509.04363 [gr-qc]].
%15 citations counted in INSPIRE as of 18 Aug 2021


%\cite{Miki:2020hvg}
\bibitem{Miki:2020hvg}
D.~Miki, A.~Matsumura and K.~Yamamoto,
%``Entanglement and decoherence of massive particles due to gravity,''
Phys. Rev. D \textbf{103}, no.2, 026017 (2021)
%doi:10.1103/PhysRevD.103.026017
[arXiv:2010.05159 [gr-qc]].
%5 citations counted in INSPIRE as of 18 Aug 2021


%\cite{Pang:2016foq}
\bibitem{Pang:2016foq}
B.~H.~Pang, Y.~Chen and F.~Y.~Khalili,
%``Universal Decoherence under Gravity: A Perspective through the Equivalence Principle,''
Phys. Rev. Lett. \textbf{117}, no.9, 090401 (2016)
%doi:10.1103/PhysRevLett.117.090401
[arXiv:1603.01984 [quant-ph]].
%24 citations counted in INSPIRE as of 18 Aug 2021


%\cite{Podolskiy:2015wna}
\bibitem{Podolskiy:2015wna}
D.~Podolskiy and R.~Lanza,
%``On decoherence in quantum gravity,''
Annalen Phys. \textbf{528}, no.9-10, 663-676 (2016)
%doi:10.1002/andp.201600011
[arXiv:1508.05377 [gr-qc]].
%4 citations counted in INSPIRE as of 18 Aug 2021

%\cite{Ellis:1988uk}
\bibitem{ellis}
J.~R.~Ellis, S.~Mohanty and D.~V.~Nanopoulos,
%``Quantum Gravity and the Collapse of the Wave Function,''
Phys. Lett. B \textbf{221}, 113-119 (1989).

\bibitem{milburn}
G.~J.~Milburn, Phys. Rev. A \textbf{44}, 5401 (1991).

\bibitem{bargueno}
P.~Bargue\~no,
%``Generalized uncertainty principle and quantum gravitational friction,''
Phys. Lett. B \textbf{727}, 496-499 (2013).

%\cite{Kempf:1996fz}
\bibitem{Kempf:1996fz}
A.~Kempf,
%``Nonpointlike particles in harmonic oscillators,''
J. Phys. A \textbf{30}, 2093-2102 (1997)
%doi:10.1088/0305-4470/30/6/030
[arXiv:hep-th/9604045 [hep-th]].
%376 citations counted in INSPIRE as of 30 Sep 2021


%\cite{Das:2017qwp}
\bibitem{Das:2017qwp}
S.~Das, M.~P.~G.~Robbins and E.~C.~Vagenas,
%``Gravitation as a source of decoherence,''
Int. J. Mod. Phys. D \textbf{27}, no.02, 1850008 (2017)
%doi:10.1142/S0218271818500086
[arXiv:1709.07154 [gr-qc]].

%\cite{Petruzziello:2020wkd}
\bibitem{Petruzziello:2020wkd}
L.~Petruzziello and F.~Illuminati,
%``Quantum gravitational decoherence from fluctuating minimal length and deformation parameter at the Planck scale,''
Nature Commun. \textbf{12}, no.1, 4449 (2021)
%doi:10.1038/s41467-021-24711-7
[arXiv:2011.01255 [gr-qc]].
%3 citations counted in INSPIRE as of 18 Aug 2021


%\cite{Ali:2010yn}
\bibitem{Ali:2010yn}
A.~F.~Ali, S.~Das and E.~C.~Vagenas,
%``The Generalized Uncertainty Principle and Quantum Gravity Phenomenology,''
%doi:10.1142/9789814374552\_0492
[arXiv:1001.2642 [hep-th]].
%46 citations counted in INSPIRE as of 18 Aug 2021


%\cite{Das:2008kaa}
\bibitem{Das:2008kaa}
S.~Das and E.~C.~Vagenas,
%``Universality of Quantum Gravity Corrections,''
Phys. Rev. Lett. \textbf{101}, 221301 (2008)
%doi:10.1103/PhysRevLett.101.221301
[arXiv:0810.5333 [hep-th]].
%448 citations counted in INSPIRE as of 18 Aug 2021


%\cite{Das:2020ujn}
\bibitem{Das:2020ujn}
A.~Das, S.~Das and E.~C.~Vagenas,
%``Discreteness of Space from GUP in Strong Gravitational Fields,''
Phys. Lett. B \textbf{809}, 135772 (2020)
%doi:10.1016/j.physletb.2020.135772
[arXiv:2006.05781 [gr-qc]].
%5 citations counted in INSPIRE as of 18 Aug 2021


%\cite{Ali:2009zq}
\bibitem{Ali:2009zq}
A.~F.~Ali, S.~Das and E.~C.~Vagenas,
%``Discreteness of Space from the Generalized Uncertainty Principle,''
Phys. Lett. B \textbf{678}, 497-499 (2009)
%doi:10.1016/j.physletb.2009.06.061
[arXiv:0906.5396 [hep-th]].
%381 citations counted in INSPIRE as of 18 Aug 2021


%\cite{Vagenas:2019wzd}
\bibitem{Vagenas:2019wzd}
E.~C.~Vagenas, A.~F.~Ali, M.~Hemeda and H.~Alshal,
%``Linear and Quadratic GUP, Liouville Theorem, Cosmological Constant, and Brick Wall Entropy,''
Eur. Phys. J. C \textbf{79}, no.5, 398 (2019)
%doi:10.1140/epjc/s10052-019-6908-z
[arXiv:1903.08494 [hep-th]].
%13 citations counted in INSPIRE as of 18 Aug 2021

%\cite{Pfister:2015sna}
\bibitem{Pfister:2015sna}
C.~Pfister, J.~Kaniewski, M.~Tomamichel, A.~Mantri, R.~Schmucker, N.~McMahon, G.~Milburn and S.~Wehner,
%``Understanding nature from experimental observations: a theory independent test for gravitational decoherence,''
Nature Commun. \textbf{7}, 3022 (2016)
[arXiv:1503.00577 [quant-ph]].
%8 citations counted in INSPIRE as of 16 Feb 2022

%\cite{Xu:2020pzr}
\bibitem{Xu:2020pzr}
B.~Xu,
%``Neutrino Decoherence in Simple Open Quantum Systems,''
[arXiv:2009.13471 [hep-ph]].
%0 citations counted in INSPIRE as of 18 Aug 2021


%\cite{Kolovsky_2020}
\bibitem{Kolovsky_2020}
A.~R.~Kolovsky
%''Quantum entanglement and the Born-Markov approximation for an open quantum system,''
Phys. Rev. E \textbf{101}, 062116 (2020)
%doi:10.1103/physreve.101.062116
[arXiv:2002.07320 [quant-ph]].


%\cite{Nakatami2010}
\bibitem{Nakatani2010}
M.~Nakatani and T.~Ogawa,
%''Quantum Master Equations for Composite Systems: Is Born–Markov Approximation Really Valid?,''
Journal of the Physical Society of Japan \textbf{79}, 8 (2010).
%doi:10.1143/JPSJ.79.084401


%\cite{Vadimov2021}
\bibitem{Vadimov2021}
V.~Vadimov, J.~Tuorila, T.~Orell, J.~Stockburger, T.~Ala-Nissila, J.~Ankerhold, and M.~M\"ott\"onen,
%''Validity of Born-Markov master equations for single- and two-qubit systems,''
Phys. Rev. B \textbf{103}, 214308 (2021)
%doi:10.1103/PhysRevB.103.214308
[arXiv:2011.05046 [quant-ph]].

%\cite{Balasubramanian:2014pba}
\bibitem{unita}
V.~Balasubramanian, S.~Das and E.~C.~Vagenas,
%``Generalized Uncertainty Principle and Self-Adjoint Operators,''
Annals Phys. \textbf{360}, 1-18 (2015)
[arXiv:1404.3962 [hep-th]].
%18 citations counted in INSPIRE as of 10 May 2023

%\cite{Dilem:2022xoj}
\bibitem{unita2}
B.~B.~Dilem, J.~C.~Fabris and J.~A.~Nogueira,
%``Self-adjoint extensions for a p4-corrected Hamiltonian of a particle on a finite interval,''
Annals Phys. \textbf{444}, 168994 (2022)
[arXiv:2204.00687 [math-ph]].
%0 citations counted in INSPIRE as of 10 May 2023

\bibitem{bose}
S.~Bose, A.~Mazumdar, G.~W.~Morley, H.~Ulbricht, M.~Toro\v{s}, M.~Paternostro, A.~Geraci, P.~Barker, M.~S.~Kim and G.~Milburn,
%``Spin Entanglement Witness for Quantum Gravity,''
Phys. Rev. Lett. \textbf{119}, no.24, 240401 (2017)
[arXiv:1707.06050 [quant-ph]].

\bibitem{vedral}
C.~Marletto and V.~Vedral,
%``Gravitationally-induced entanglement between two massive particles is sufficient evidence of quantum effects in gravity,''
Phys. Rev. Lett. \textbf{119}, no.24, 240402 (2017)
[arXiv:1707.06036 [quant-ph]].










%\cite{Lindblad:1975ef}
%\bibitem{Lindblad:1975ef}
%G.~Lindblad,
%``On the Generators of Quantum Dynamical Semigroups,''
%Commun. Math. Phys. \textbf{48}, 119 (1976).
%doi:10.1007/BF01608499
%832 citations counted in INSPIRE as of 18 Aug 2021


%\cite{Zurek:1982ii}
%\bibitem{Zurek:1982ii}
%W.~H.~Zurek,
%``Environment induced superselection rules,''
%Phys. Rev. D \textbf{26}, 1862-1880 (1982).
%doi:10.1103/PhysRevD.26.1862
%443 citations counted in INSPIRE as of 18 Aug 2021


%\cite{Lidar:1998hs}
%\bibitem{Lidar:1998hs}
%D.~A.~Lidar, I.~L.~Chuang and K.~B.~Whaley,
%``Decoherence free subspaces for quantum computation,''
%Phys. Rev. Lett. \textbf{81}, 2594 (1998)
%doi:10.1103/PhysRevLett.81.2594
%[arXiv:quant-ph/9807004 [quant-ph]].
%81 citations counted in INSPIRE as of 18 Aug 2021


%\cite{Joos:1984uk}
%\bibitem{Joos:1984uk}
%E.~Joos and H.~D.~Zeh,
%``The Emergence of classical properties through interaction with the environment,''
%Z. Phys. B \textbf{59}, 223-243 (1985).
%doi:10.1007/BF01725541
%461 citations counted in INSPIRE as of 18 Aug 2021


%\cite{Breuer:2008rh}
%\bibitem{Breuer:2008rh}
%H.~P.~Breuer, E.~Goklu and C.~Lammerzahl,
%``Metric fluctuations and decoherence,''
%Class. Quant. Grav. \textbf{26}, 105012 (2009)
%doi:10.1088/0264-9381/26/10/105012
%[arXiv:0812.0420 [gr-qc]].
%30 citations counted in INSPIRE as of 18 Aug 2021


%\cite{Pikovski:2013qwa}
%\bibitem{Pikovski:2013qwa}
%I.~Pikovski, M.~Zych, F.~Costa and C.~Brukner,
%``Universal decoherence due to gravitational time dilation,''
%Nature Phys. \textbf{11}, 668-672 (2015)
%doi:10.1038/nphys3366
%[arXiv:1311.1095 [quant-ph]].
%101 citations counted in INSPIRE as of 18 Aug 2021


%\cite{Scardigli:1999jh}
%\bibitem{Scardigli:1999jh}
%F.~Scardigli,
%``Generalized uncertainty principle in quantum gravity from micro - black hole Gedanken experiment,''
%Phys. Lett. B \textbf{452}, 39-44 (1999)
%doi:10.1016/S0370-2693(99)00167-7
%[arXiv:hep-th/9904025 [hep-th]].
%564 citations counted in INSPIRE as of 18 Aug 2021



%\cite{Amati:1987wq}
%\bibitem{Amati:1987wq}
%D.~Amati, M.~Ciafaloni and G.~Veneziano,
%``Superstring Collisions at Planckian Energies,''
%Phys. Lett. B \textbf{197}, 81 (1987)
%doi:10.1016/0370-2693(87)90346-7
%712 citations counted in INSPIRE as of 18 Aug 2021


%\cite{Blencowe:2012mp}
%\bibitem{Blencowe:2012mp}
%M.~P.~Blencowe,
%``Effective Field Theory Approach to Gravitationally Induced Decoherence,''
%Phys. Rev. Lett. \textbf{111}, no.2, 021302 (2013)
%doi:10.1103/PhysRevLett.111.021302
%[arXiv:1211.4751 [quant-ph]].
%69 citations counted in INSPIRE as of 18 Aug 2021










\end{thebibliography}
\end{document}